# An analysis of the abstracts presented at the annual meetings of the Society for Neuroscience from 2001 to 2006

John M Lin[1], Jason W Bohland[1], Peter Andrews[1], Gully Burns[2], Cara B Allen[1], Partha P Mitra[1]

[1]Cold Spring Harbor Laboratory, Cold Spring Harbor, NY, USA
[2] Information Sciences Institute, University of Southern California, Los Angeles, CA, USA

## ABSTRACT

Continuing exponential growth in the volume of science as measured by number of scientists or by publications has made it virtually impossible for individual researchers to keep track of the totality of knowledge and major progress areas in a research field using the traditional modes of scholarly research. This is individually frustrating for researchers not satisfied with exploring increasingly hyper-specialized niches, but also has negative implications for broader questions relating to the efficiency of the research enterprise and for science policy. Automated or semi-automated methods using natural language processing, applied to the scientific literature, provide a potential avenue to address this problem. Indeed, such bibliometric analysis forms the groundwork for search engines such as Google. However, most of the scientific literature is behind a Byzantine arrangement of online firewalls which prevent efficient utilization of automated tools by the average researcher. Meeting abstracts published by scientific societies are often available freely in electronic form on the web or in form of media distributed at annual meetings, and forms an attractive starting point for the construction and mining of knowledge bases about specific scientific domains. In particular, the annual meeting of the Society for Neuroscience (SFN) is a large-scale, international event that is arguably the most influential single meeting within the subject. The abstracts of presentation at this meeting are not peer-reviewed publications, but nonetheless provide a unique global survey of the state of the subject of neuroscience each year.

In this paper, we extracted and processed abstract data from the SFN annual meeting abstracts during the period 2001-2006, using a suitable combination of relatively standard software ranging from natural language processing and database management to visualization and analysis tools. An important first step in the process was the application of data cleaning and disambiguation methods to construct a unified database, since the data were too noisy to be of full utility in the raw form initially available. The resulting co-author graph in 2006, for example, had 39,645 nodes (with an estimated 6% error rate in our disambiguation of similar author names) and 13,979 abstracts, with an average of 1.5 abstracts per author, 4.3 authors per abstract, and 5.96 collaborators per author (including all authors on shared abstracts). Most authors (28,084) have a single abstract; 6665, 2464, and 1002 authors have 2, 3, and 4 abstracts, respectively.

Recent work in related areas has focused on reputational indices such as highly cited papers or scientists and journal impact factors, and to a lesser extent on creating visual maps of the knowledge space. In contrast, there has been relatively less work on the demographics and community structure, the dynamics of the field over time to examine major research trends and the structure of the sources of research funding. In this paper we examined each of these areas in order to gain an objective overview of contemporary neuroscience including its demographics and community structure, major research areas and trends, and the distribution of NIH funding across topic clusters. Some interesting findings include a high geographical concentration of neuroscience research in north eastern United States, a surprisingly large transient population (60% of the authors appear in only one out of the six studied years), the central role played by the study of neurodegenerative disorders in the neuroscience community structure, and an apparent growth of behavioral/systems neuroscience with a corresponding shrinkage of cellular/molecular neuroscience over the six year period.

## RESULT AND DISCUSSION

We extracted and processed data from the annual Society for Neuroscience (SFN) meeting planners to build databases of SFN abstracts and their authors. Maintaining an accurate count of the total number of authors was a challenging task complicated by two types of ambiguities: (1) different authors may share the same name and initials, and (2) the same author may use different number of initials in different abstracts. In this study, we used a combination of string matching, entity matching, and co-authorship patterns to disambiguate unique authors. See Materials and Methods for details of these processes. We created one database for each year between 2001 and 2006, as well as a consolidated database encompassing data from all 6 years. The information contained in these databases allowed us to perform a variety of analyses to elucidate the structure and evolution of the neuroscience landscape.

The rest of the paper is organized as follows. First, we present the geographical distribution of the SFN authors, followed by basic statistics and demographics of the SFN annual meetings. We constructed a graph of co-authors on abstracts and applied graph theoretic algorithms to investigate patterns of connection and communication between neuroscientists. Finally, we used computational techniques in natural language processing to cluster the abstracts into neuroscience topics and studied their dynamics and concordance of these discovered topic clusters with the thematic organization provided by the SFN. We also studied the distribution of NIH funding across these topics.

### 1. Geographical Distribution of SFN Abstract Authors

To explore geographical distribution and dynamics of neuroscience research, the city, state (for US and Canada), and country of each author's institution was extracted. The number of authors associated with each unique location was then tabulated for each year between 2001 and 2006. Table 1 shows the top 10 cities with the highest SFN representation during this time frame. Based on these data, the global "hubs" for neuroscience research seem to be concentrated in the following geographical regions: northeast region of the United States (Boston, New York, Philadelphia, Baltimore/DC vicinity), Southern California, Tokyo, Montreal, and London. These representations remained fairly static over the years, indicating the stable presence of prominent and well-funded neuroscience research centers in these regions. By plotting the changes in the percentage of representation for some of these locations (Figure 1), it is evident that New York City consistently ranks as the top producer of neuroscience research, signifying the number and caliber of academic institutions, research centers, and hospitals in the New York metropolitan area. In addition, the city of Atlanta appears to have a steadily increasing presence in the neuroscience landscape, although the spike occurred between the year 2005 and 2006 may be partly attributed to the fact that the 2006 SFN meeting was held in that city.

It is interesting to compare this list with the top ten cities in terms of scientific publications in 1967 (Table 7.2 in Price, 1986). In descending order, these were Moscow, London, New York, Paris, Tokyo, Washington, Boston, Philadelphia, Chicago, St Petersburg (Leningrad).

| 2001 | 2002 | 2003 | 2004 | 2005 | 2006 |
|---|---|---|---|---|---|
| New York | New York | New York | New York | New York | New York |
| Boston | Baltimore | Bethesda | Los Angeles | Baltimore | Baltimore |
| Baltimore | Bethesda | Baltimore | Boston | Bethesda | Boston |
| Los Angeles | Boston | Boston | Bethesda | Boston | Bethesda |
| Bethesda | Los Angeles | Los Angeles | La Jolla | Los Angeles | Los Angeles |
| La Jolla | Tokyo | La Jolla | Baltimore | Philadelphia | Chicago |
| Tokyo | Chicago | Chicago | Philadelphia | Chicago | Atlanta |
| London | Philadelphia | Philadelphia | Chicago | La Jolla | Philadelphia |
| Montreal | La Jolla | London | Tokyo | Atlanta | Tokyo |
| Chicago | Montreal | Tokyo | Pittsburgh | Tokyo | La Jolla |

**Table 1.** Top 10 cities for SFN representation between 2001 and 2006

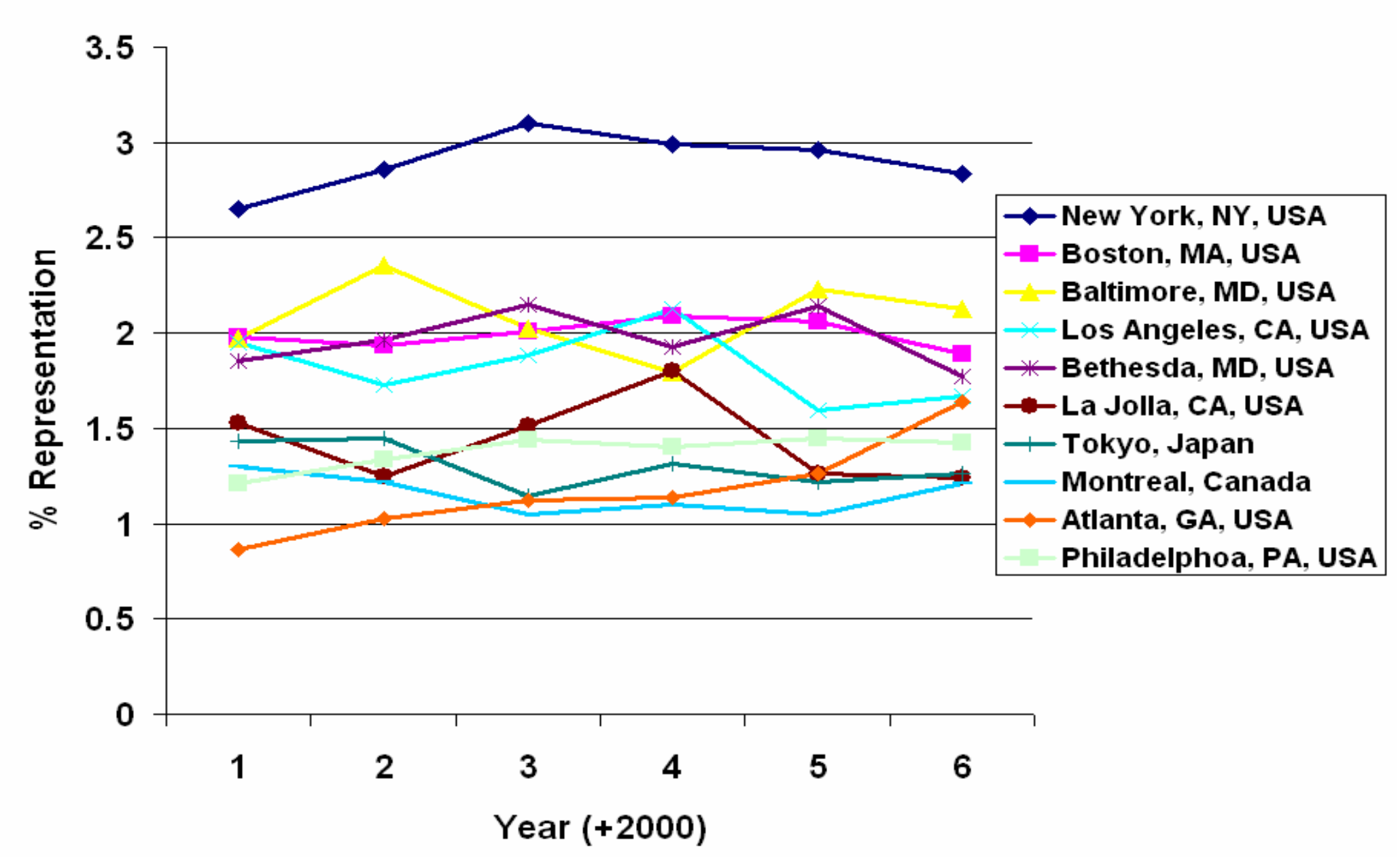

**Figure 1.** Changes in the percentage of representation for several geographical locations between the year 2001 and 2006

The advent of web mapping technologies such as GoogleMap (maps.google.com), YahooMap (maps.yahoo.com) provides capabilities to generate, visualize, and navigate high quality geographical maps on the World Wide Web. In order to visualize the geographical distribution of the home institutions of abstract authors on a map, the latitude and longitude of each address from the abstracts were fetched using Yahoo's GeoCode Web Service (http://developer.yahoo.com/maps/rest/V1/geocode.html). The quantitative distribution of these geographical data can then be plotted on different map templates using the application programming interface (API) provided by the mapping engine. Figure 2 shows different perspectives of geographical distribution of 2006 SFN data using Google Map API.

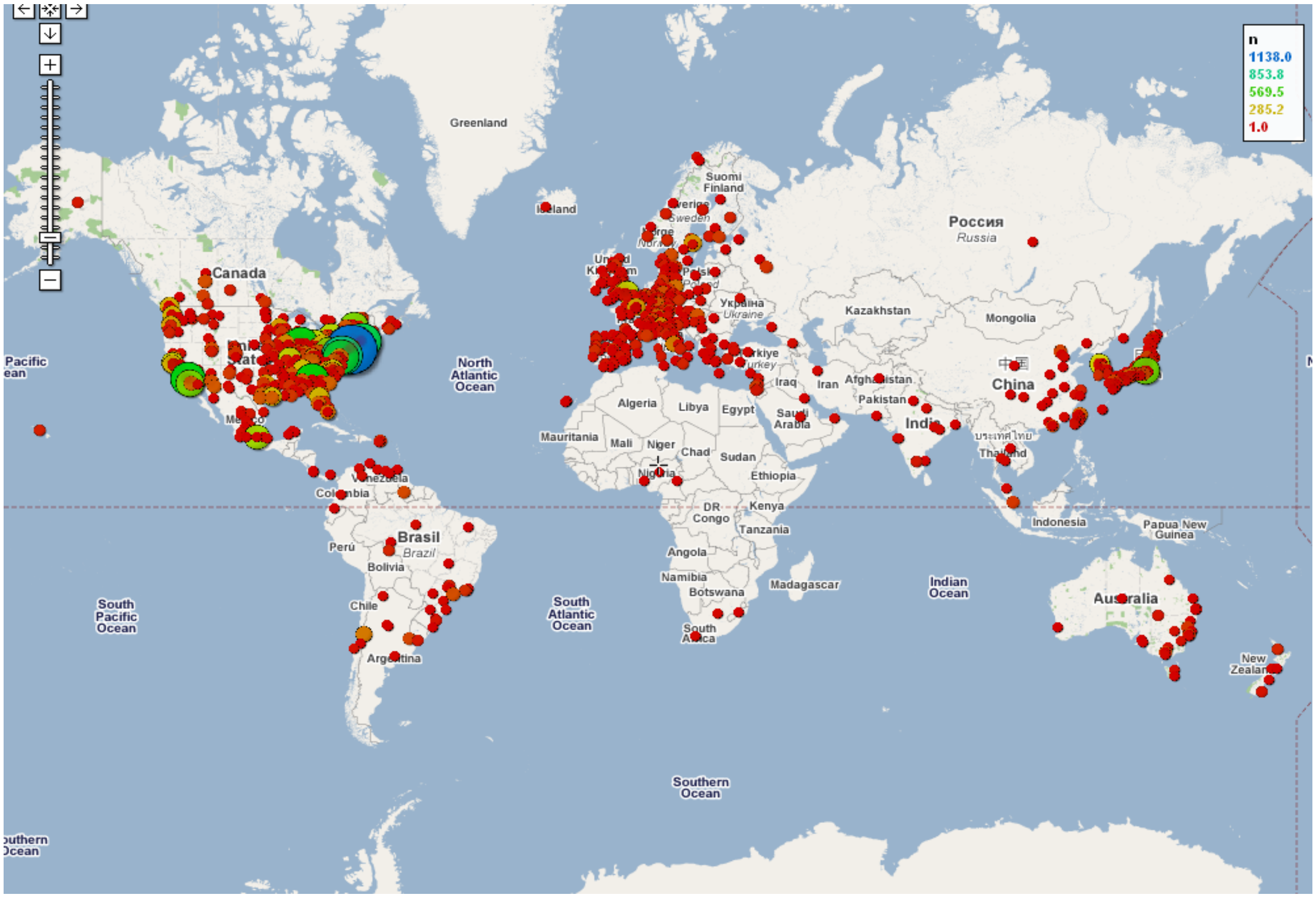

**Figure 2.** Geographical distribution of SFN authors for the year 2006 displayed on a Google Map. The location with the most representation is in the northeast region of the US, as indicated by the blue circle. Other regions of high representation include Southern California (Los Angeles and San Diego), London, and Tokyo.

## 2. Basic Statistics and Demographics

The upper bound for the total number of authors in the six-year database is 197429; this number was obtained by parsing the data from SFN abstracts without applying any of the disambiguation or entity matching schemes described in the Materials and Methods section. After these schemes were applied, the total number of authors was reduced by approximately 35% to 128553. The number of unique author names in the database was 99410, which represents the lower bound for total author count. The final tally of 128553 falls between the upper and lower bound and gives a reasonable estimate of the true number of unique authors in the database.

The top five countries represented in the SFN annual meetings between 2001 and 2006 were: USA (56.6%), EU-15 (20.2%), Japan (7.3%), Canada (5.2%), and Mexico (1.4%). For comparison, the share of worldwide science and engineering article production in 2003 was 31.5% (EU-15), 30.3 (USA) and 8.6 (Japan) (National Science Board, 2006). Note also that the number of life scientists employed in the science and technology workforce in the USA in 2000 was estimated to be 226,000 (National Science Board, 2006). As discussed below, most (60%) of the SFN abstract authors appear to be transients.

For the 6-year data between 2001 and 2006, the average number of abstracts per author was 2.93, and the average number of authors per abstract was 4.31. Looking at the statistics on a year by year basis (Table 2), it is apparent that the number of abstracts per author, number of authors per abstract, and average number of collaborators in any given year stay roughly constant during the six year span. This suggests that the neuroscience community produces research results at a relatively constant rate and that most research projects in the field are conducted by a small to moderate team of scientists. The average number of authors on Science and Engineering articles worldwide in 2003 was reported to be 4.22 and the corresponding number for the US was 4.42 (National Science Board, 2006), suggesting that the team sizes represented in SFN abstracts are consistent with other areas of science.

| Year | Number of Authors | Number of Abstracts | Avg Abstracts Per Author | Avg Authors Per Abstract | Average Number of Collaborators |
|---|---|---|---|---|---|
| **2001** | 42318 | 15340 | 1.55 | 4.28 | 5.82 |
| **2002** | 37129 | 13307 | 1.53 | 4.21 | 5.51 |
| **2003** | 41349 | 15261 | 1.58 | 4.29 | 5.90 |
| **2004** | 43853 | 15987 | 1.59 | 4.37 | 6.09 |
| **2005** | 39622 | 13669 | 1.50 | 4.35 | 5.88 |
| **2006** | 39645 | 13979 | 1.54 | 4.33 | 5.96 |
| **2001-2006** | 128553 | 87543 | 2.93 | 4.31 | 8.62 |

**Table 2.** Basic statistics of SFN data for the 6-year period between 2001 and 2006

To further elucidate the collaboration patterns of neuroscientists, we plotted the histograms of the number of co-authors for abstracts and the number of abstracts submitted by authors. As highlighted in Figure 3(a), most SFN meeting abstracts contain two to five authors. Very few abstracts are associated with only one author or more than 10 authors. This may again imply that most research projects in neuroscience are carried out by a few scientists instead of large teams of people.

Figure 3(b) shows the histograms of the number of abstracts associated with each author. The majority of the authors had only one abstract over the span of six years. We speculated that this number reflected a large group of "transients" comprising mostly undergraduates, graduate students, and perhaps post-docs who entered and exited the neuroscience field in a short period of time. Given the increasingly blurred boundaries between different disciplines of biomedical sciences, it is possible that many of these scientists simply shifted their focus to a different aspect of biomedical research, i.e. from cellular neuroscience to genomics, or from cognitive neuroscience to psychology. The histograms also highlight a few individuals who are associated with very large numbers of abstracts (some have over 100).

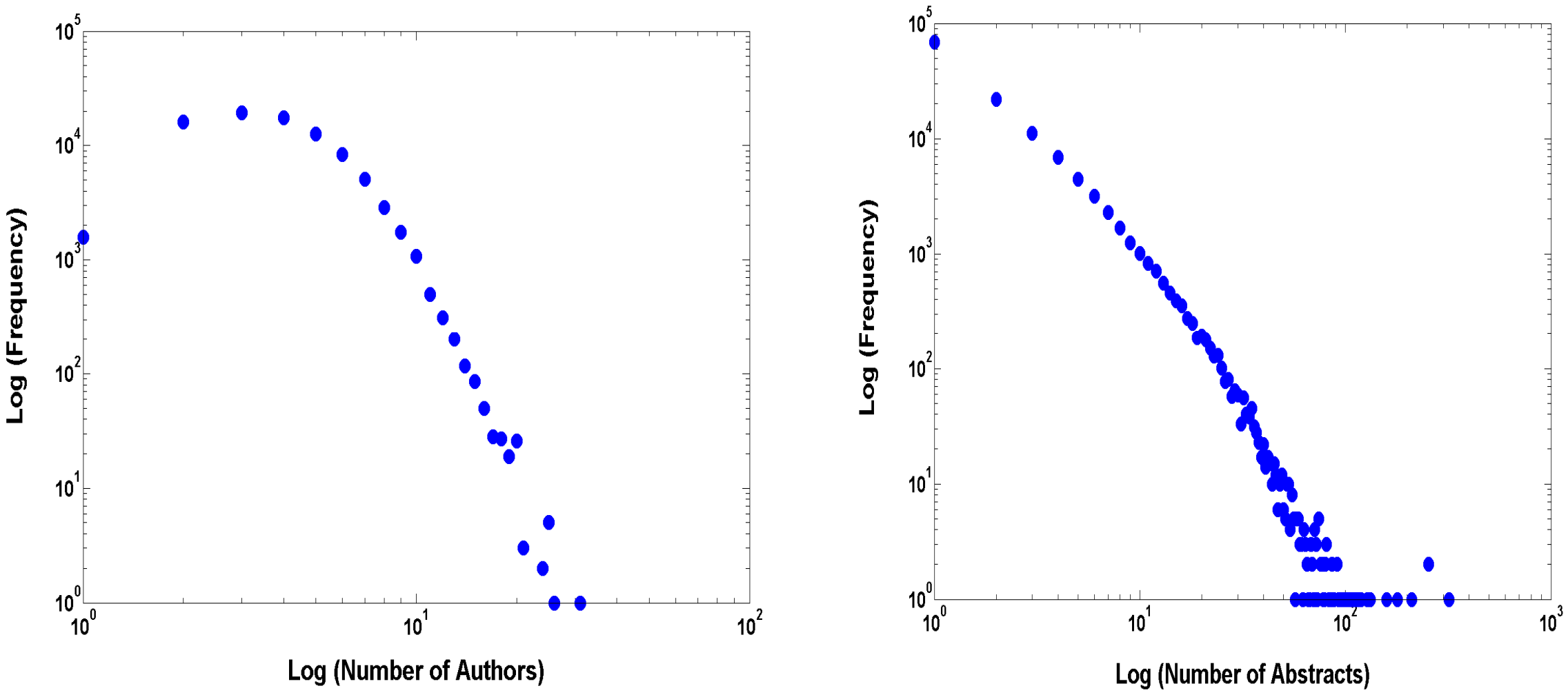

(a) (b)

**Figure 3.** (a) Histograms of the average number of authors per abstract between 2001 and 2006. (b) Histograms of the average number of abstracts per author between 2001 and 2006.

In Figure 4(a), we plotted the histograms of the number of years in which authors are represented between 2001 and 2006. As the figure shows, approximately 60% of the authors made presence in only one SFN meeting within the six-year period. Again, we speculated that this high turn over rate is the direct manifestation of many transients who entered and exited the field in a relatively short time frame. The phenomena of a high transient rate, reflecting a sort of "infant mortality rate" for first time authors was first analyzed by Price (Price, 1986), who estimated a 22% transient rate for paper authorship from a database consisting of a statistical sample of papers published between 1964 and 1970. Although we do not pursue it in detail, it should be straightforward to extend or implement Price's model of transients and continuants to the SFN abstracts database, particularly if data from a longer period of time becomes available.

To correlate these data with the demographics of actual SFN meeting attendance, we downloaded from the SFN website (www.sfn.org) the annual meeting attendance statistics from 1971 to 2006. These data are plotted in Figure 4(b) using a base 2 logarithm. The SFN meeting attendance has shown an overall slowing growth rate in the past 3 decades. As evident from the graph, the first doubling took approximately 5 years. The next two doublings occurred at a steady exponential rate between 1975 and 1995, with a doubling period of about 8 years. The growth slowed after 1995 and the current doubling rate is projected to be about 15 years.

What are the causes of the exponential growth, and what is causing the rates to slow down? To put the numbers in perspective, the number of life scientists employed in the Science and Technology workforce in the US for the years 1970, 1980, 1990 and 2000 were 55, 102, 139 and 226 (in thousands). These numbers also show an initial doubling period of 10 years and a subsequent slowing. The exponential increase in the number of scientists and scientific publication over the last three centuries has been studied systematically (Price, 1986).

Interestingly, Price's estimate of the doubling times of 10-15 years is consistent with the estimated growth of SFN meeting attendees over the last three decades. However, this growth has also slowed down and may continue to fall further. It is interesting to speculate about what is slowing the growth in meeting attendance. A number of limitations come to mind: a reflection of an overall slowdown in growth of the S&T workforce or in general or of biomedical scientists in particular, perhaps due to saturating funding rates; maturation of the research field and a shift in scientific talent to other growth areas such as information technology, or perhaps non-scientific factors such as the number of hotel rooms in the cities where the meetings are held,

To see if the participation level of the annual SFN meetings might be linked to the amount of funding available to the scientists, we plotted in Figure 4(c) the budgets for National Institute of Health (NIH), one of the largest funding agencies for biomedical sciences, from 1976 to 2006 (Source: Historical Table 2: R&D by agency, AAAS website http://www.aaas.org/spp/rd/guihist.htm). Although NIH budget has grown steadily during this period, there does not seem to be a detailed correlation between NIH funding and SFN meeting attendance. In fact, the growth in meeting attendance slowed down precisely when NIH budget was doubled from 1995 to 2005.

Exponential growths do not continue forever, and the increase in the number of SFN attendees is no exception. Price has pointed out that a doubling time of 10-15 years is much faster than the doubling time of the human population (which is currently around 50 years, and slowing), and has predicted a period of transition to a steady state where the number of scientists per capita reaches a stable value. In Price's estimate, we are either at the inflection point of the corresponding logistic curve, or have passed it already. It is to be noted that the percentage of GNP devoted to R&D in developed nations has remained steady between 2-3% since the 70's (Ziman, 1990), and other subject areas in science such as physics or electrical engineering also showed sharp growth followed by saturation within recent history. Unfortunately, despite such historical data and exhortations by Price and others about the necessity to manage the transition from rapid exponential growth to slower growth or a relatively steady state, there is little evidence for forward planning by the biomedical community in trying to manage the coming demographic transition by practicing stricter scientific "birth control" (Martinson, 2007). Absent such planning, the danger is that Malthusian factors will make the transition significantly more painful than necessary.

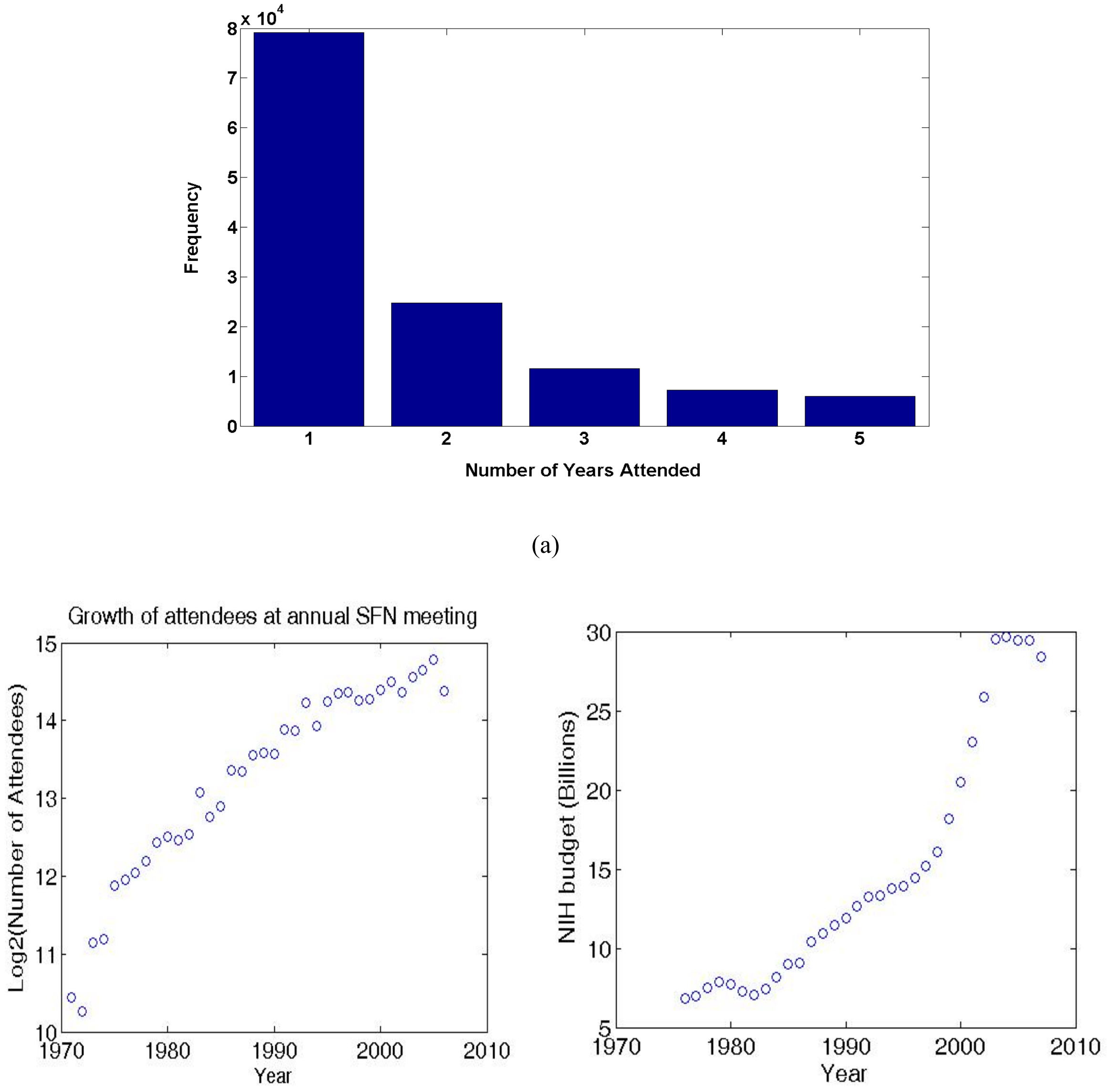

(a)

(b) (c)

**Figure 4.** (a) Histograms of the number of years in which authors are represented. (b) Growth of attendees at annual SFN meeting from 1971 to 2006 (in base 2 logarithm). (c) Growth of NIH funding from 1976 to 2006.

## 3. Analysis of Co-authorship Graphs

A co-author graph, $G: = (V, E)$, was constructed from the preprocessed database by representing each author as a vertex on a graph, $v \in V$. Two authors were connected by an undirected edge, $e \in E$, if they have co-authored at least one abstract in the database. Matrix representations of the graph can be used to analyze the structure of the underlying community. In addition, by integrating the data with a graph visualization package, such as Graphviz (www.graphviz.org) or JUNG (jung.sourceforge.net) one can visualize, explore, and navigate the network interactively.

A fundamental measure used in graph theory is the shortest path between a pair of connected vertices. In the context of the network under study, this measures the number of steps it takes to go from one author to another through intermediate collaborators. From the multi-year SFN database, the lengths of shortest paths between all pairs of authors for whom a connection exists were calculated exhaustively using breadth-first search algorithm. These numbers were then averaged to yield the mean distance between authors in the entire network. Table 3 shows that the authors in the SFN community are separated from one another by an average distance of 6.09. A similar observation of "six degree of separation" has been reported previously for abstracts in the MEDLINE database (Newman, 2001), suggesting that neuroscience and the greater biomedical science community share similar connection patterns. The diameter of the graph, or the maximum distance between any two authors in the network for whom a connection exists, is 20, which also closely matches the result from Newman's MEDLINE analysis.

We also computed the clustering coefficient, which measures cliquishness (Watts and Strogatz, 1998). Suppose that a vertex $v$ in a graph has $k_v$ neighbors; then at most $k_v(k_v-1)/2$ edges can exist between them (this occurs when every neighbor of $v$ is connected to every other neighbor of $v$). Let $C_v$ denote the fraction of possible edges for the neighborhood around $v$ that actually exist. The clustering coefficient of a graph is the average of $C_v$ for all $v$. The mean clustering coefficient for the SFN network between 2001 and 2006 is 0.7724. In other words, two authors in the network have a 77.24% or greater probability of being collaborators if they have both collaborated with a third author.

A large sparse graph such as the one created from the SFN database may not be *connected* (i.e. there may not exist a path from each vertex to every other vertex in the graph). Finding the set of individual connected components in the graph may provide another insight into community structure. The SFN coauthor graph for 2001-2006 was found to contain 2650 connected components (Table 3). Most authors belong to a single large connected component which comprises more than 90% of the entire network. The remaining connected components in the graph are significantly smaller, each accounting for less than 1% of the vertices of the entire graph. Some of these small connected components represent research groups from pharmaceutical companies or other commercial entities, while some others belong to laboratories from countries with a relatively low SFN presence.

Another interesting aspect of the graph is the relative importance of each vertex as measured by the *betweenness centrality* of the vertex (Anthonisse, 1971; Freeman, 1977). The betweenness centrality for a given vertex $BC(v)$ is defined as:

$$BC(v) = \sum_{s \neq v \neq t \in V} \frac{\sigma_{st}(v)}{\sigma_{st}}$$

where $\sigma_{st}$ is the number of shortest paths between $s \in V$ and $t \in V$, and $\sigma_{st}(v)$ is the number of shortest paths between $s$ and $t$ that pass through $v$. In other words, betweenness centrality measures the frequency with which a vertex falls on one of the shortest paths between any other pair of vertices in the graph.

Vertices with large betweenness have more influence over the information flow in the graph and can thus be considered to represent authors playing central roles in the SFN co-author network. Analysis of the multi-year SFN data revealed that only a few individuals in the network have disproportionately large betweenness centrality measures (Figure 5(a)). In addition, Figure 5(b) shows that on average the distribution of the betweenness centrality of an author and the number of abstracts closely follow a power law. However, the authors possessing the largest betweenness centrality, and thus the most influence over the network, were not necessarily associated with the most number of abstracts. To better elucidate the roles of these brokering members of the SFN network, the research profiles of these individuals were located from the World Wide Web and qualitatively assessed. Most of the authors with high betweenness centrality conduct research in the field of neurodegenerative diseases, such as Alzheimer's disease (AD) and Parkinson's disease (PD). Research related to AD, PD, and other neurodegenerative diseases is highly multidisciplinary in nature, and scientists engaging in this type of research will likely employ techniques and methodologies spanning multiple different sub-disciplines of neuroscience and other biomedical sciences, which might explain the high values of betweenness centrality. Another possible reason is the comparatively high funding rates for neurodegenerative disorders (discussed in a later section).

| | |
|---|---|
| Average Distance | 6.09 |
| Graph Diameter | 20 |
| Mean Clustering Coefficient | 0.7724 |
| Number of Connected Components | 2650 |
| Size of Largest Connected Component | 116716 |
| As a percentage | 90.79% |
| Size of Second Largest Connected Component | 56 |
| As a percentage | 0.0436% |

**Table 3.** Some graph analysis results for multi-year SFN data

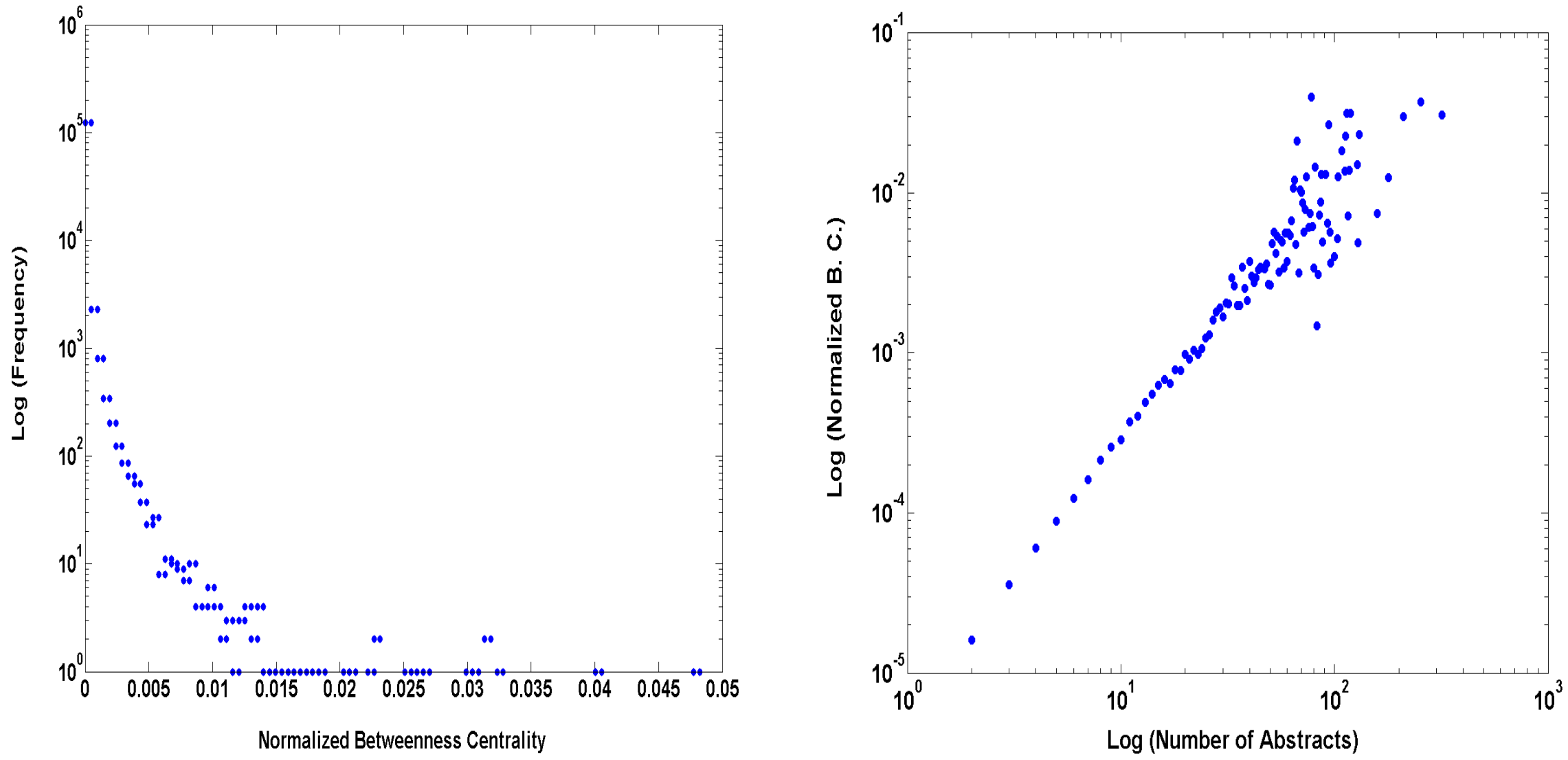

(a) (b)

**Figure 5.** (a) Histograms of the betweenness centrality (bc) normalized by total number of possible edges, [N*(N-1)]/2, where *N* is the number of authors, from all authors plotted in log scale. The majority of the authors have very small normalized bc (less than 0.005), and only a few authors have disproportionately large bc. (b) The averaged normalized bc over all authors having the same number of abstracts as a function of the number of abstracts. On average, the betweenness centrality of an author and the number of the abstracts follow a power law.

## 4. Topic Modeling

The sheer number and diversity of the annual SFN meeting attendees indicate that the text corpora from the abstracts provide an illustrative view of the current state and dynamics of the neuroscience research landscape. One can perform a variety of text mining and natural language processing (NLP) techniques to exploit topic information from the syntaxes and semantics of the text corpora. The information gained from topic modeling can be used to classify abstracts into different categories, chart the rise and fall of research topics over time, measure the popularity of specific fields, and facilitate document retrieval.

In this work, we explored the utility of *Latent Semantic Analysis* (LSA), (Deerwester et al., 1990), to find the topic space spanned by the SFN abstract set. Briefly, LSA is a dimensionality reduction technique that projects terms and documents (abstracts) to a lower dimensional space. The reduced dimensionality vector space captures most of the important underlying structure in the association of terms and documents, while at the same time removing the noise or variability in word usage (Berry et al., 1995). In the reduced vector space, terms that occur in similar documents are located near one another even if they never co-occur in the same document, and topically related documents are grouped near one another based on their semantic relatedness.

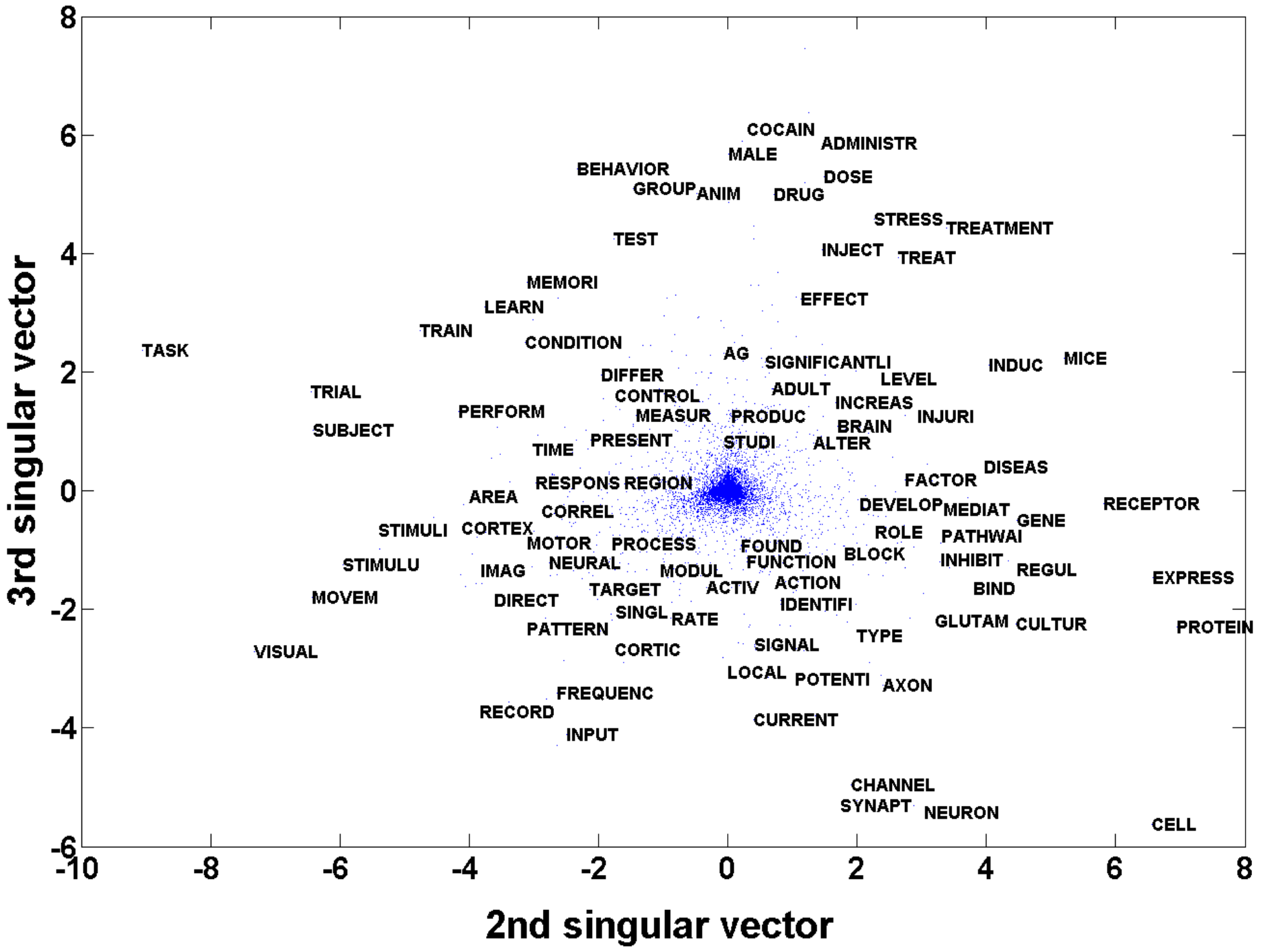

**Figure 6.** Projections of the terms (represented by the blue dots) on the reduced vector space formed by the 2nd and 3rd singular vectors of the truncated Singular Value Decomposition (See Materials and Methods). Select terms with high frequencies are labeled in the figure. Note that these terms were stemmed.

Figure 6 shows the projections of the terms in a reduced two-dimensional vector space. The terms with the highest frequencies of occurrence are labeled. It can be seen from the figure that some terms representing similar concepts are located near one another in this reduced vector space. For example, many terms on the left side of the figure are related to sensory and motor systems (“task”, “stimulus”, “movement”, “visual”), terms at the bottom of the figure are related to cellular neuroscience (“potential”, “current”, “axon”, “channel”, “synaptic”, “neuron”), and many terms on the right side of the figure are related to molecular biology (“protein”, “gene”, “regulatory”, “bind”, “express”, “pathway”).

After LSA was performed using 100 dimensions, we applied the *Normalized Cuts* (NCuts) algorithm (Shi and Malik, 1997) to automatically cluster the abstracts into different topic groups. The number of topic clusters was determined by evaluating concordance between the topic labels found by clustering to the eight SFN theme labels (Figure 7) which were available in the database. The concordance evaluation was performed using the *Adjusted Rand Index* (Hubert and Arabie, 1985), a measure that quantifies the agreement between two clustering systems. The number of topic clusters that maximized concordance was found to be 10 (Refer to the Materials and Methods section for detailed descriptions of these algorithms).

**Theme A**: Development
**Theme B**: Neural Excitability, Synapses, and Glia: Cellular Mechanisms
**Theme C**: Sensory and Motor Systems
**Theme D**: Homeostatic and Neuroendocrine Systems
**Theme E**: Cognition and Behavior
**Theme F**: Disorders of the Nervous System
**Theme G**: Techniques in Neuroscience
**Theme H**: History and Teaching of Neuroscience

**Figure 7**. Themes used by SFN to categorize abstracts submitted for the 2006 meeting

## 4.1 Topic Clusters

To understand the content of the resulting topic clusters, we found the 20 most frequent words used in each cluster. The lists of frequent words, along with the complete collections of the abstracts, were also distributed to laboratory members working in neuroscience for subjective labeling. Among the 10 topic clusters, half of them were readily identified for their distinct and coherent themes. For example, all abstracts in Cluster 3 deal with research in songbirds. Abstracts in Cluster 6 frequently contain such words as “amyloid beta”, “abeta”, “tau protein”, or other terms relevant to Alzheimer’s disease. Cluster 7 is distinct from all other clusters in that it contains mostly education and informatics related work. Cluster 8 groups together all abstracts related to biological rhythms, which is evident from the abundance of the following words: “circadian”, “melatonin”, “clock”, “phase”, and “suprachiasmatic nucleus” or “SCN”. Finally, Cluster 10 contains mostly abstracts dealing with the structures and mechanisms of sleep. The remaining 5 clusters, which tend to be larger in size, were not as readily identifiable and required more thorough investigation of the abstracts themselves. Table 4 shows cluster sizes, lists of frequent words, and the labels qualitatively assigned to each cluster. For illustrative purpose, only the 7 most distinguishing words taken from each cluster’s list of 20 most frequent words are shown. Complete lists of the 20 most frequent words for each cluster are available as supplementary materials.

To visualize the 10 topic clusters on a high level “conceptual map”, the abstracts from all six years were plotted as points in a 2D space formed by the two leading eigenvectors of the graph Laplacian defined on the nearest neighbor abstracts graph. Each abstract was color coded based on the topic cluster to which it belongs. The resulting maps are presented in Figure 8.

| 1 (16729) | 2 (22647) | 3 (492) | 4 (7210) | 5 (19988) |
|---|---|---|---|---|
| *Substance Abuse & Addiction* | *Cellular Neuroscience* | *Behavior of Song Birds* | *Pain & Trauma* | *Proteins, Gene Expression & Molecular Biology* |
| BEHAVIOR<br>LEVEL<br>COCAINE<br>DOSE<br>DRUG<br>INJECT<br>TREATMENT | SYNAPTIC<br>PROTEIN<br>CURRENT<br>CHANNEL<br>POTENTIAL<br>DENDRITIC<br>SUBUNIT | SONG<br>HVC*<br>BIRD<br>VOCAL<br>AUDITORY<br>FUNCH<br>SING | SPINAL<br>PAIN<br>RECEPTOR<br>MUSCLE<br>INJURIES<br>DORSAL<br>MORPHINE | CELL<br>NEURON<br>EXPRESS<br>ACTIVE<br>BRAIN<br>GENE<br>RECEPTOR |

| 6 (3609) | 7 (736) | 8 (794) | 9 (14192) | 10 (1146) |
|---|---|---|---|---|
| *Alzheimer's Disease* | *Education & Informatics* | *Biological Rhythms* | *Visual & Motor Systems* | *Sleep* |
| AD*<br>AMYLOID<br>TAU<br>ALZHEIMER<br>PEPTIDE<br>PLAQUE<br>ABETA | STUDENT<br>DATA<br>LEARN<br>PROGRAM<br>MODEL<br>SCHOOL<br>INFORMATICS | CIRCADIAN<br>SCN*<br>LIGHT<br>RHYTHM<br>PHASE<br>CLOCK<br>CYCLE | RESPONSE<br>TASK<br>VISUAL<br>SUBJECT<br>CORTEX<br>MOVEMENT<br>STIMULUS | SLEEP<br>WAKE<br>REM*<br>EEG<br>DEPRIVATION<br>PERIOD<br>WAVE |

*Abbreviations: HVC = "High Vocal Center"; AD = "Alzheimer's Disease"; REM = "Rapid Eye Movement"; SCN = "Suprachiasmatic Nucleus"

**Table 4.** The 10 clusters produced by the Ncuts algorithm performed on the nearest-neighbor graph from year 2001-2006 (see Materials and Methods). The size of each cluster is given in parentheses. The second row of the table shows the subjective assessed topic assigned by domain experts to each cluster. The third row shows the 7 most distinguishing words found in the 20 most frequently used words in each cluster.

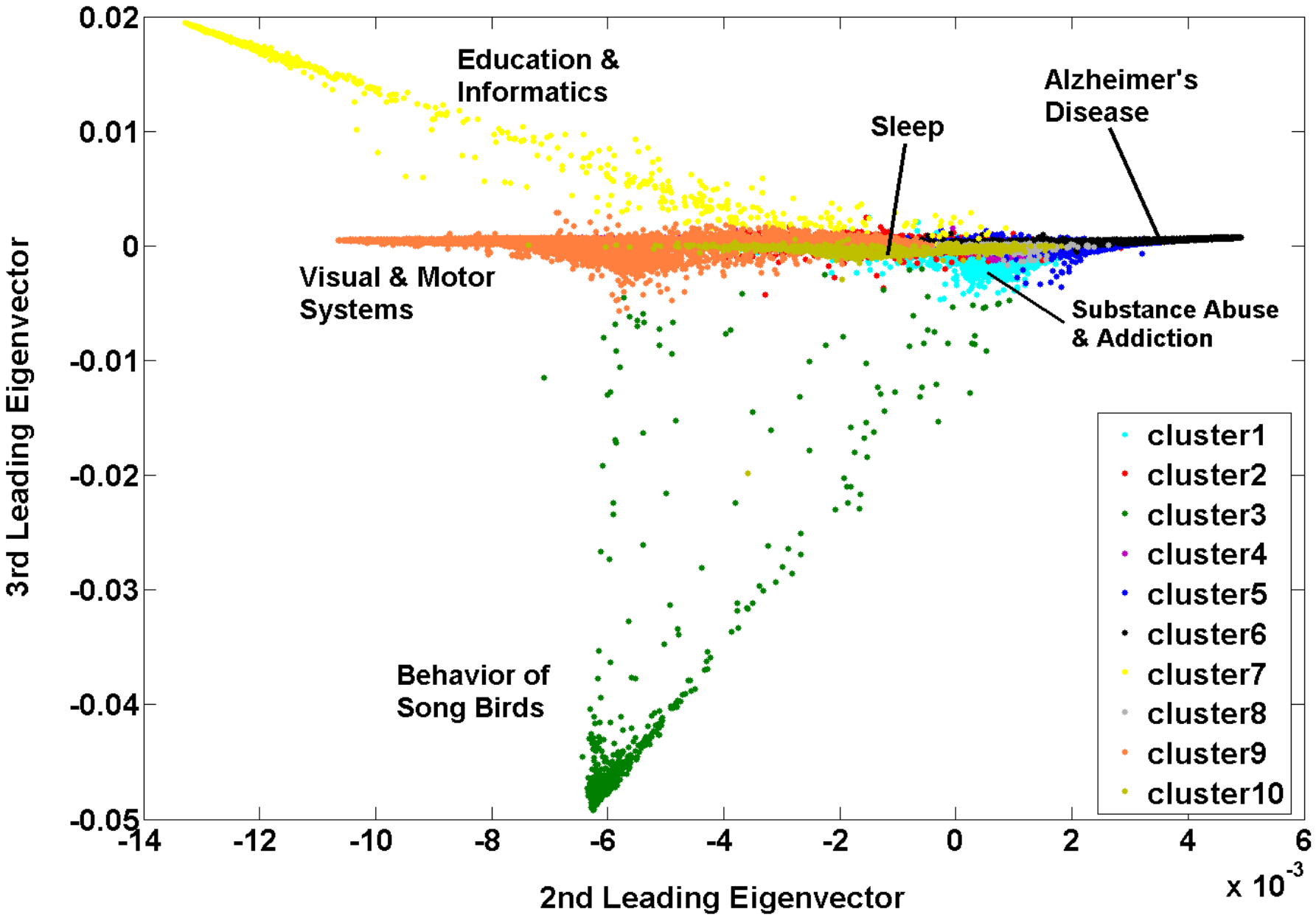

(a)

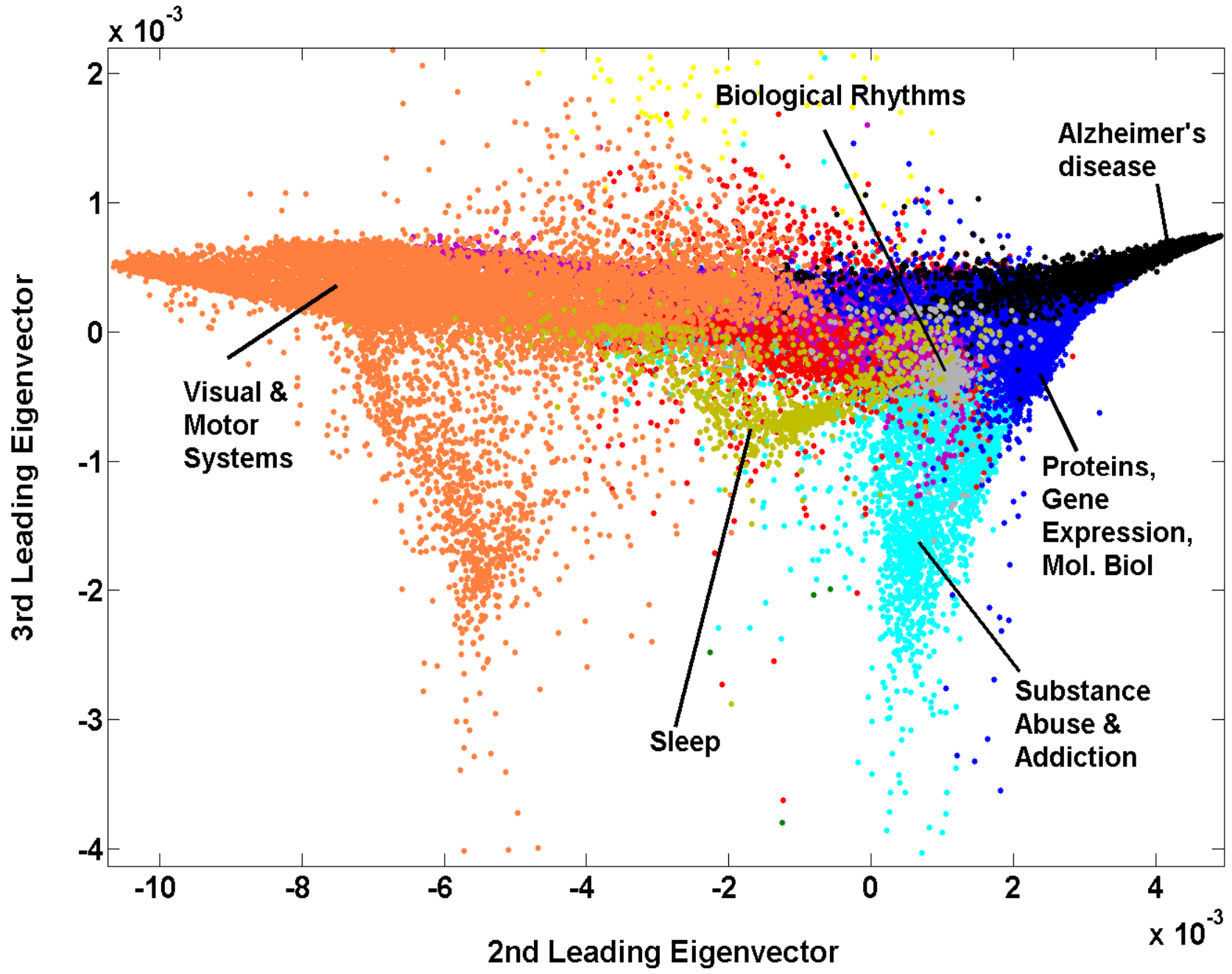

(b)

**Figure 8.** (a) Visualization of topic map for all SFN meeting abstracts from 2001 to 2006. Abstracts assigned to different clusters appear in different colors (see legend). (b) Zooming in at the center of the topic map reveals more detailed clusters

## 4.2 Concordance with SFN Themes

While the Adjusted Rand Index provides a global measure of similarity between partitions, the individual abstract clusters derived from NCuts partitioning can be compared pairwise with the SFN theme clusters. A concordance matrix, *C*, between the two classification systems was constructed in which each element $C_{ij}$ indicates the number of abstracts from the 2006 that belonged to cluster *i* and theme *j* (i=[1..10], j=[A..H]).

Figure 9(a) shows the relative distribution of abstracts in each cluster across the SFN themes, after dividing each matrix element $C_{ij}$ by the total number of abstracts in cluster *i*. Thus these matrix entries represent the proportion of abstracts from cluster *i* that are classified as theme *j*. Some observations of good concordance can be made:

- Most of the abstracts from Cluster 7 ("Education and Informatics") are labeled as Theme G (Techniques in Neuroscience) or Theme H (History and Teaching of Neuroscience), with more percentage in the latter.
- Cluster 6, which represents Alzheimer's disease, is almost wholly contained in Theme F (Disorders of the Nervous System).
- Cluster 3, which corresponds to behavior of song birds, is mostly captured by Theme E (Cognition and Behavior).
- There is fairly good concordance between Cluster 4, which represents topics related to pain and trauma, and Theme C (Sensory and Motor Systems).
- Good concordance is also observed between Cluster 8 ("Biological Rhythms") and Theme D (Homeostatic and Neuroendocrine Systems).

Similarly, by dividing each matrix element $C_{ij}$ by the total number of abstracts in theme *j*, the resulting matrix (Figure 9(b)) gives the proportion of abstracts from theme *j* that are classified as cluster *i*. There are some interesting observations as well:

- Theme H (History and Teaching of Neuroscience) is almost entirely contained in Cluster 7.
- Theme G (Techniques in Neuroscience) is spread between Cluster 2 ("Cellular Neuroscience") and Cluster 9 ("Visual and Motor Systems"). This illustrates that while SFN groups together techniques used in kinematics, imaging, and cellular neuroscience, unsupervised clustering classified these abstracts according to their target applications.
- There is very good concordance between Theme B (Neural Excitability, Synapses, and Glia: Cellular Mechanisms) and Cluster 2.
- Many abstracts from Theme D belong to Cluster 1 ("Substance Abuse and Addiction"), suggesting that mechanisms of addiction to various psychoactive substances (i.e. alcohol, tobacco, drugs) are important elements of homeostatic and neuroendocrine research.

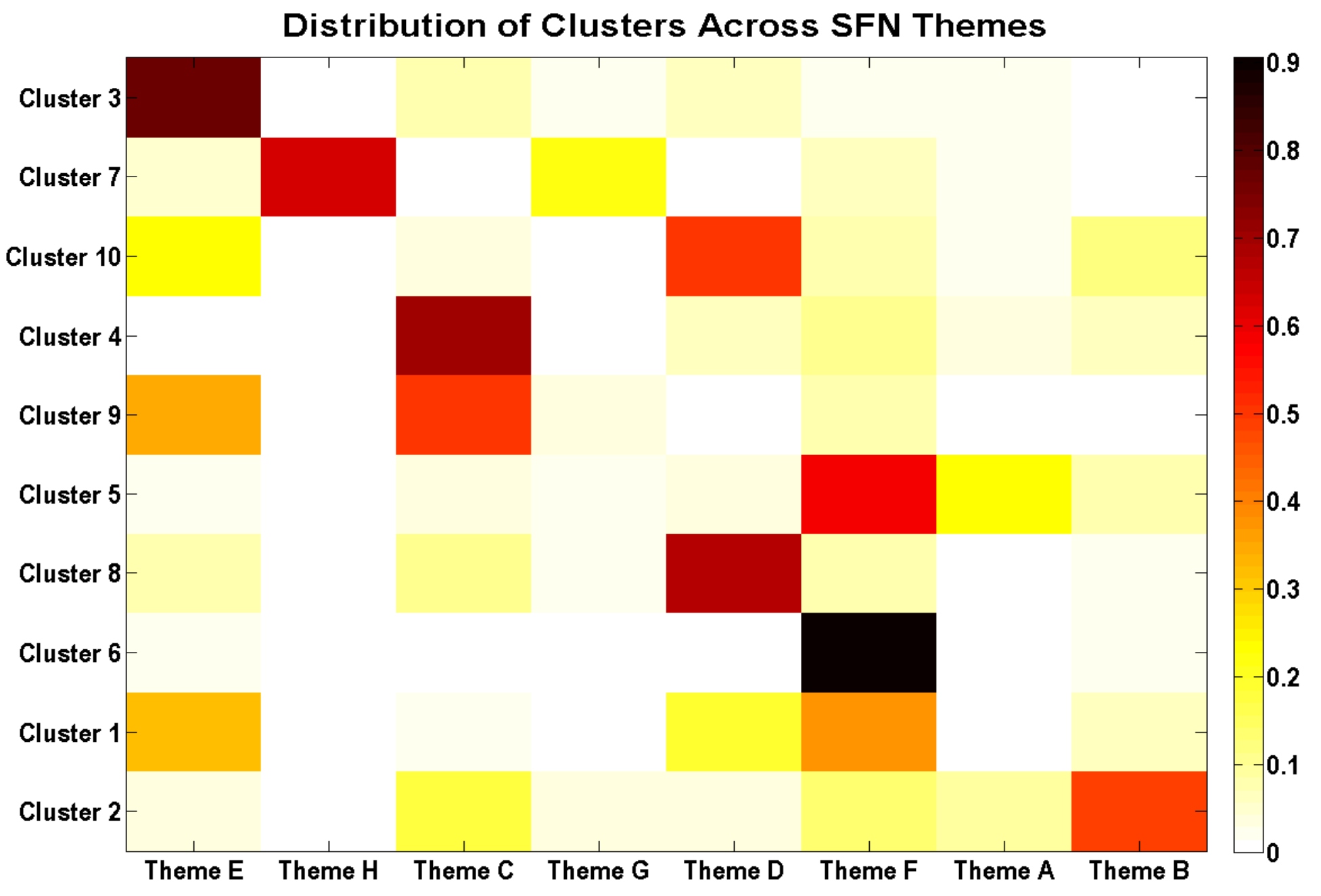

(a)

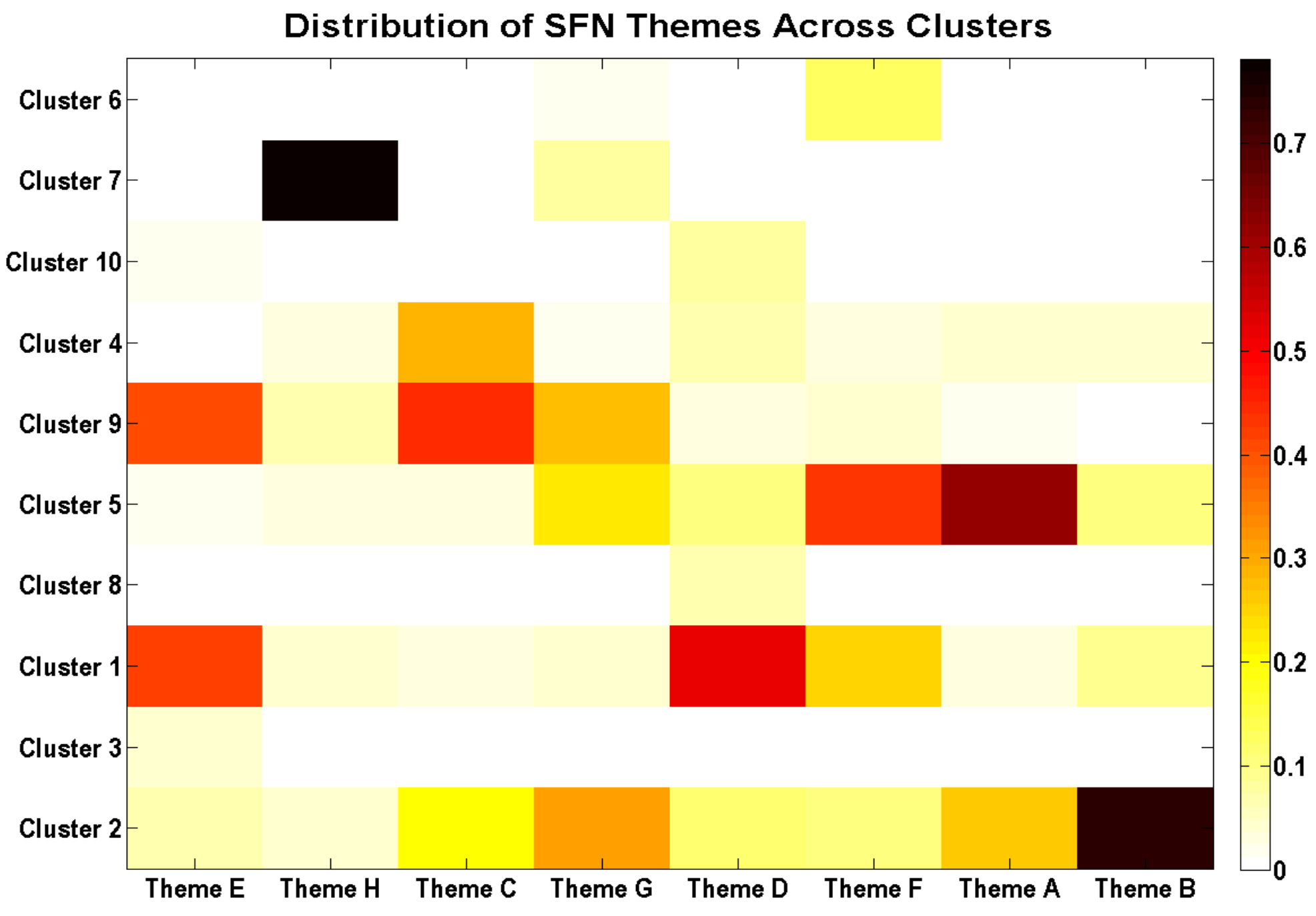

(b)

**Figure 9.** (a) Concordance matrix between Ncuts clusters and SFN themes, normalized by cluster size in each row. The matrix has been diagonalized for clarity. (b) Concordance matrix between Ncuts clusters and SFN themes, normalized by theme size in each column. The matrix has been diagonalized for clarity.

### 4.3 Dynamics of Topics

Analyzing the dynamics of scientific topics provides interesting insights into the rise and fall of different research subjects and methodologies. The amount of scientific interest generated by different topics has both sociological and economical implications, and tracking their changes can potentially prove useful for policy making, research planning, and funding allocation. Since the topic clustering performed in the previous section was applied to a corpus of abstracts spanning 6 years, it is straightforward to study short-term trends in neuroscience research by examining how the distribution of abstracts across the topic clusters changes from year to year. Detailed descriptions of this method are outlined in Materials and Methods.

Among the 10 topic clusters, Cluster 9, which corresponds to visual and motor systems, is shown to consistently increase in representation over the six year span (Figure 10(a)). On the other hand, Cluster 2, which corresponds to cellular neuroscience, exhibits the most significant decrease in representation over the same period (Figure 10(b)). These results suggest that there is a shift in general scientific interest from cellular-level work such as ion channel, synapse, and membrane physiology, towards more system level research incorporating such topics as vision, kinematics, motor processing, and imaging. We speculated this trend is reflective of the heavy reliance of neuroscience research on animal models and invasive techniques. The use of animal model systems continues to be the most prevalent way of studying the pathophysiologic mechanisms of neurodegenerative diseases, which is an area which is both well funded and well represented in the SFN abstracts database. This may explain the rise of macro-level study in favor of cell-based and molecular techniques. In addition, neuroimaging technologies have in recent years become indispensable tools in various aspects of neuroscience research. It is therefore not surprising to observe a surge of activities related to this subject matter.

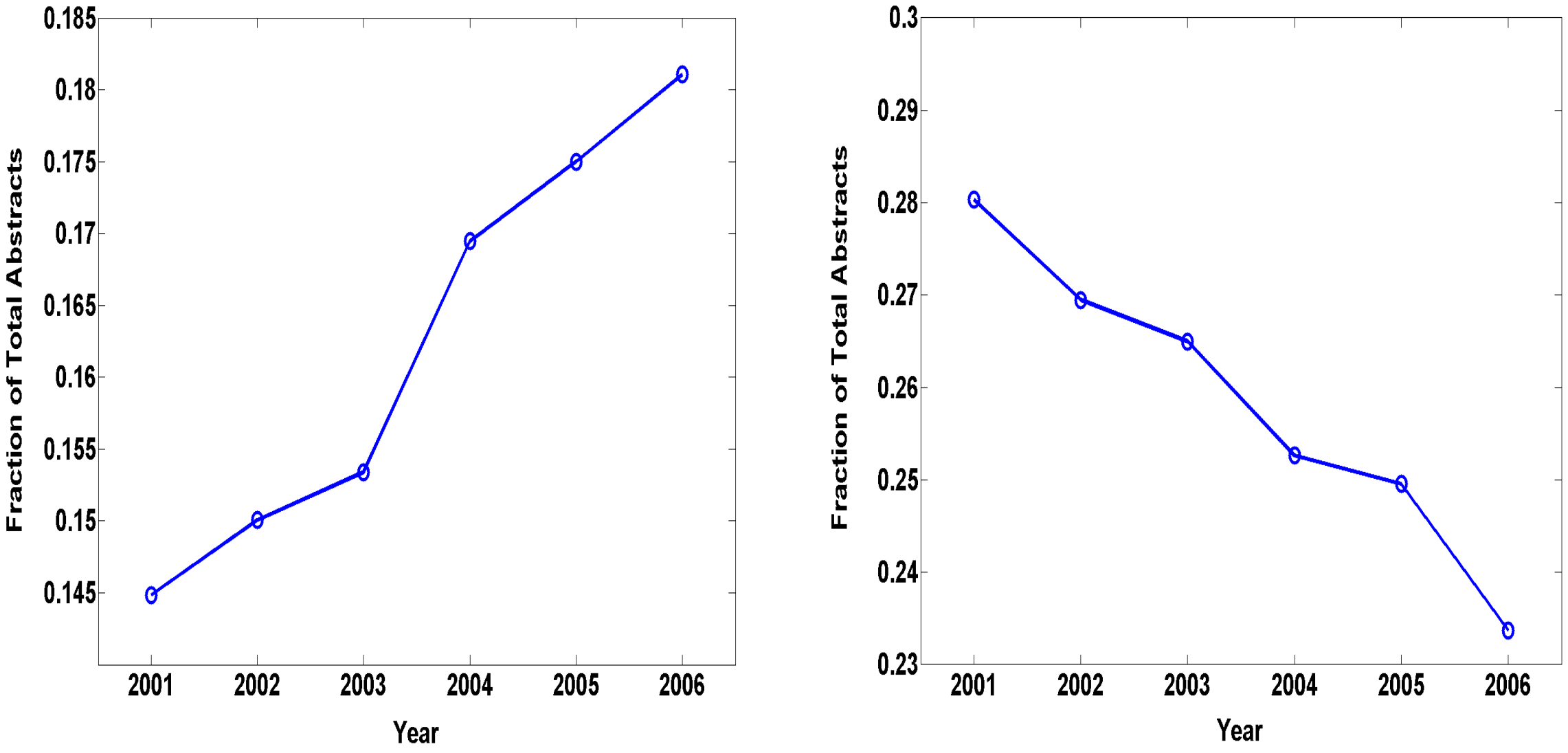

(a) (b)

**Figure 10.** (a) Dynamics of Cluster 9 ("Visual and Motor Systems"), which shows consistent and strong increase in representation from 2001 to 2006. (b) Dynamics of Cluster 2 ("Cellular Neuroscience"), which shows steady decrease in representation from 2001 to 2006.

In addition to charting the changes in the distribution of abstracts across topic clusters, we also performed analysis of word frequency dynamics using principal component analysis (see Materials and Methods). The results revealed that the first principal component accounted for over 74% of the variance in $F$ (Figure 11, top left). The first temporal component, which is roughly linear across time, is shown in Figure 11 (top). The corresponding first singular vector in word-space was sorted in order to find the specific terms that have the largest (positive and negative) projections on this temporal series. The words with largest negative projections (bottom left) are decreasing in frequency, whereas the words with the largest positive projections (bottom right) are increasing in frequency.

The results of principal component analysis on word-frequency dynamics indicated that a large fraction of the changes could be accounted for by a nearly linear component in time, which intuitively corresponds simply to some words becoming more frequent and some becoming less frequent. The corresponding word-space vector was examined to see which words contributed to the increase and which to the decrease. Figure 11 (bottom) shows the 25 terms with the largest positive and negative projections on this component. These terms seem to roughly correspond to the domains of *cellular neuroscience* (decreasing) and *systems neuroscience* (increasing). This finding is consistent with the analysis of topic clustering dynamics (above), and appears to indicate a significant shift in the topics being addressed at the Society for Neuroscience conference between the years 2001 and 2006.

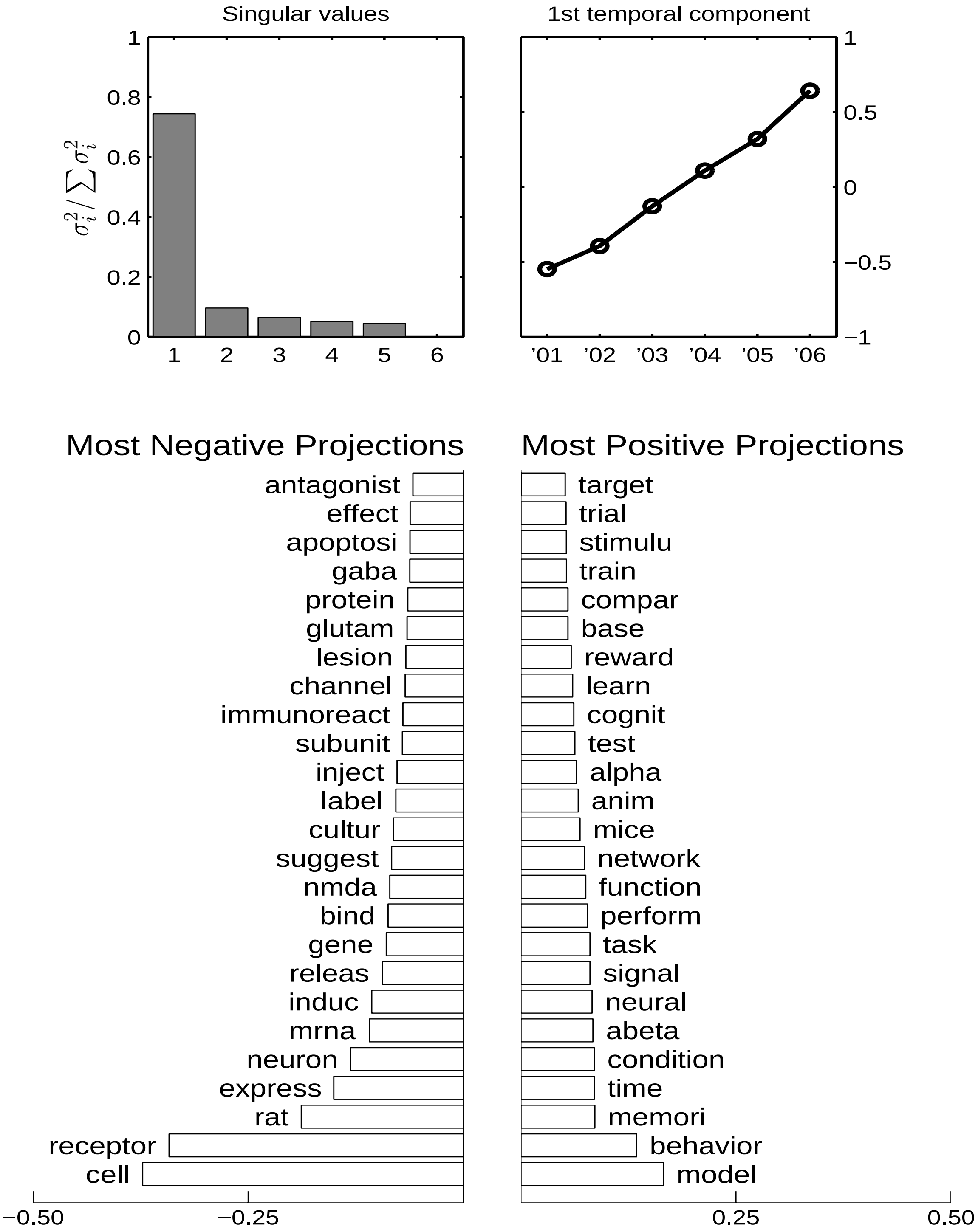

**Figure 11.** Word-frequency dynamics. Top left: distribution of singular values. The first component accounts for 74.4% of variance. Top right: The first right singular vector (temporal component). Bottom: The most negative and most positive projections of specific words onto the first component. Most positive words are increasing in frequency; most negative words are decreasing.

### 4.4 NIH Funding Analysis

The National Institutes of Health (NIH) is the largest funding agency for biomedical research in the world, currently investing over $28 billion each year for conducting and supporting medical research in the US and around the world (from NIH website: http://www.nih.gov/about/budget.htm). The NIH is made up of 27 different institutes, each of which manages research activities related to specific topics. The list of these 27 institutes is provided in Table 5. Much of the research showcased in SFN meetings is supported completely or partially by the NIH institutes. The correspondence between research dollars allocated from individual NIH institute and topic clusters provides another interesting perspective of the current neuroscience landscape. As a caveat to this section, it should be noted that the derivation of the funding information from the abstracts is inferential, since no dollar figures are provided in the abstracts, and we did not make any attempt to find tune our analysis to individual funding mechanisms but counted each listed grant equally. Nevertheless, no comparably comprehensive database of neuroscience funding is publicly available, and we considered it valuable to perform such inferential analysis.

We anticipated a correspondence between certain topic clusters and specific NIH institutes. For example, Figure 12(a) shows the NIH funding breakdown among the 8 themes created by SFN for the 2006 meeting abstracts. The majority of the work categorized as Theme A ("Disorders of the Nervous System) was supported by the National Institute of Neurological Disorders and Stroke (NINDS) and National Institute of Drug Abuse (NIDA). If we further explore the funding distributions among the subthemes of Theme F (Figure 12(b)), it is clear that neurodegenerative disorders and addiction and drugs of abuse indeed represent the majority of the work classified as Theme F. Applying the same analysis to the NCuts topic clusters, one might expect to find many abstracts from Cluster 1 (subjectively labeled "Substance abuse and addiction") to be supported by the National Institute of Drug Abuse (NIDA), and most of the work supported by the National Eye Institute (NEI) to be captured by Cluster 9 ("Visual and Motor Systems").

The funding information associated with each abstract between 2001 and 2006 was parsed from the original XML data file. If the NIH was designated as one of the funding sources, the specific institute was determined from the two-letter organization code preceding the grant number. For abstracts supported by more than one grant, an appropriate fraction was assigned to each institute by dividing the number of grants from each institute by the total number of grants listed. It should be pointed out that not all abstracts provided support information, and not all of those that did provided a grant number. However, considering the size of the database, the result is likely to be representative of the overall funding breakdown among the institutes. The breakdown of funding across the topics derived from NCuts and the NIH institutes is illustrated in Figure 12(c).

As an example of an inference that may be drawn from these visualizations, note that a large fraction of neuroscience research, both at the cellular and system level, is supported by NINDS. This observation is consistent with the expectation that, regardless of techniques or methodologies, one of the ultimate goals of many neuroscience investigations is to further the understanding of the causes, prevention, diagnostics, and treatment of various disorders of the nervous systems. If more detailed information can be extracted from the specific grants referenced, one might further break down NINDS funding among different types of neurological disorders. These types of information can be useful for research planning and analysis of the societal costs of neurological diseases.

There is good concordance between several NIH institutes and our topic clusters. For example, most abstracts from Cluster 6, which corresponds to Alzheimer's disease (AD), are supported by the National Institute on Aging (NIA). Similarly, a significant portion of the abstracts funded by the NIA are from Cluster 6, suggesting that AD is indeed the top neurological health priority for the aging population. As anticipated, another example of good concordance is the fact that most of the work supported by NEI is associated with Cluster 9, which encompasses visual and motor systems. Finally, it makes intuitive sense that NIDA and National Institute on Alcohol Abuse and Alcoholism (NIAAA) would apportion most resources to support works related to substance abuse and addiction, which is captured by Cluster 1.

| Acronym | Full Name | Organizational Code |
|---|---|---|
| CLC | Clinical Center | CL |
| CSR | Center for Scientific Review | RG |
| FIC | John E. Fogarty International Center | TW |
| NCCAM | Natl. Ctr. for Complementary and Alternative Medicine | AT |
| NCI | National Cancer Institute | CA |
| NCMHD | National Center on Minority Health and Health Disparities | MD |
| NCRR | National Center for Research Resources | RR |
| NEI | National Eye Institute | EY |
| NHGRI | National Human Genome Research Institute | HG |
| NHLBI | National Heart, Lung, and Blood Institute | HL |
| NIA | National Institute on Aging | AG |
| NIAAA | National Institute on Alcohol Abuse and Alcoholism | AA |
| NIAID | National Institute of Allergy and Infectious Diseases | AI |
| NIAMS | Natl. Inst. of Arthritis and Musculoskeletal & Skin Diseases | AR |
| NIBIB | Natl. Inst. of Biomedical Imaging and Bioengineering | EB |
| NICHD | National Institute of Child Health and Human Development | HD |
| NIDA | National Institute on Drug Abuse | DA |
| NIDCD | Natl. Inst. on Deafness and Other Communication Disorders | DC |
| NIDCR | National Institute of Dental and Craniofacial Research | DE |
| NIDDK | Natl. Inst. of Diabetes and Digestive and Kidney Diseases | DK |
| NIEHS | National Institute of Environmental Health Sciences | ES |
| NIGMS | National Institute of General Medical Sciences | GM |
| NIMH | National Institute of Mental Health | MH |
| NINDS | National Institute of Neurological Disorders and Stroke | NS |
| NINR | National Institute of Nursing Research | NR |
| NLM | National Library of Medicine | LM |
| OD | Office of the Director | OD |

**Table 5.** List of NIH Institutes, acronyms, and organizational codes (extracted from the website http://grants.nih.gov/grants/glossary.htm).

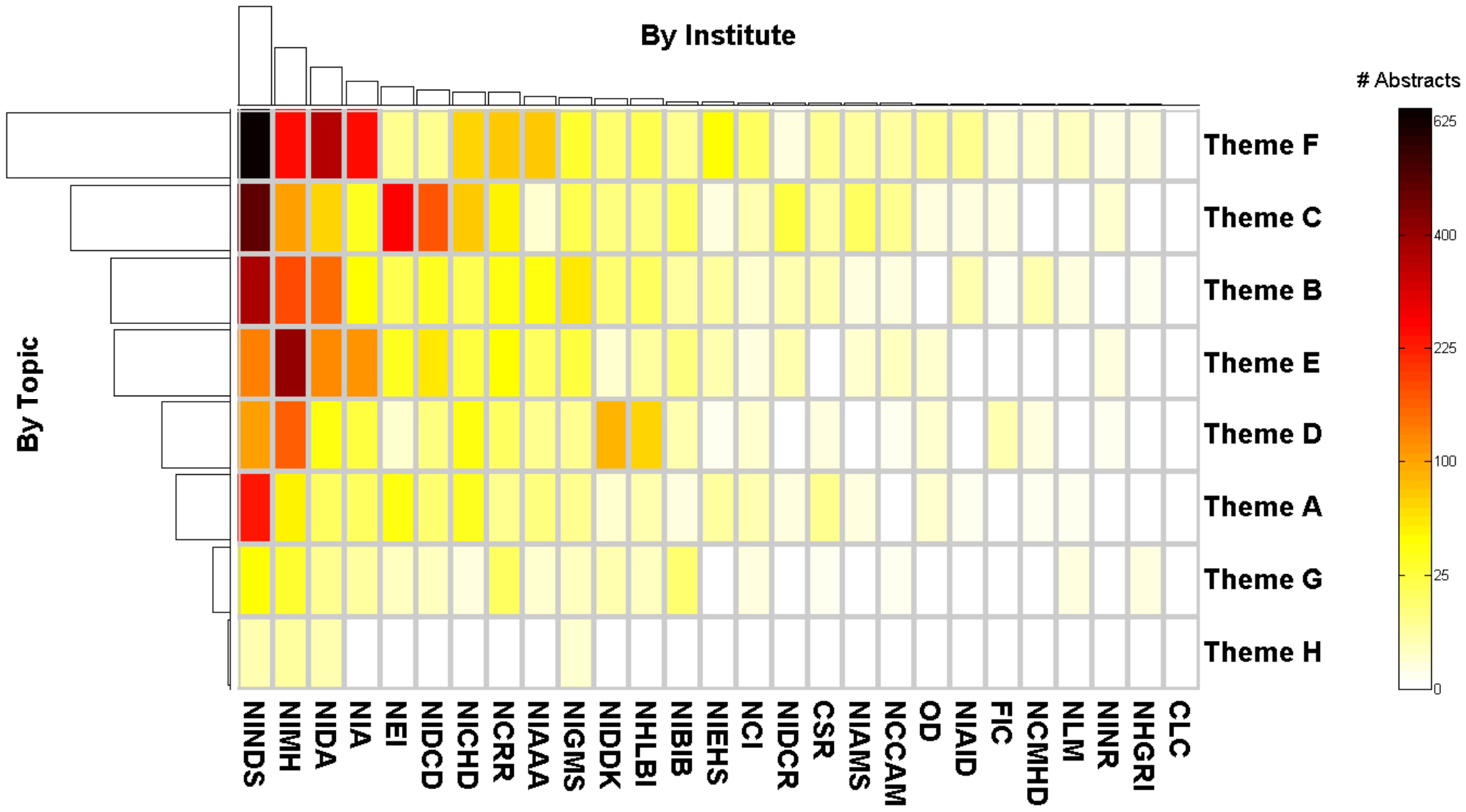

(a)

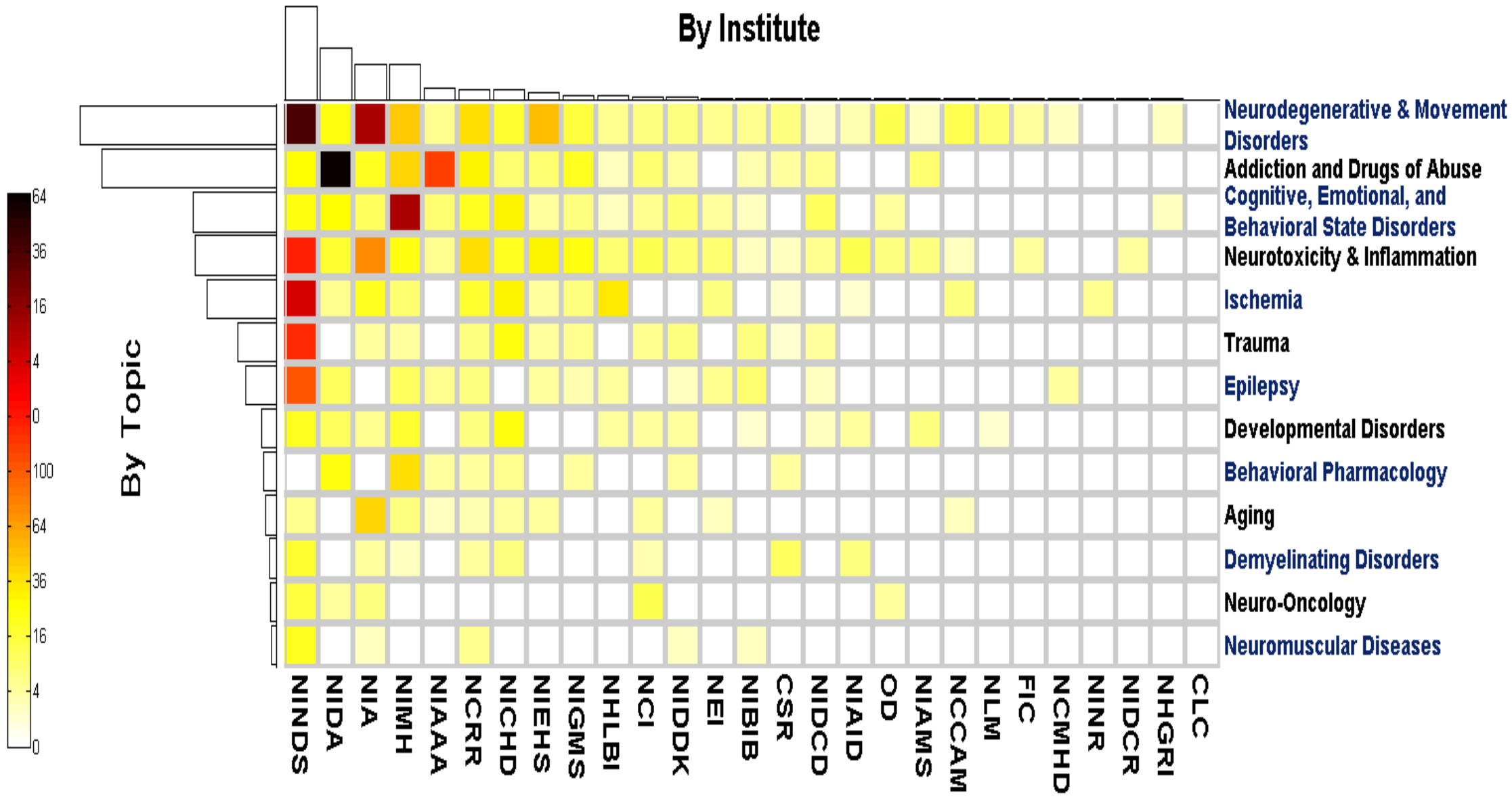

(b)

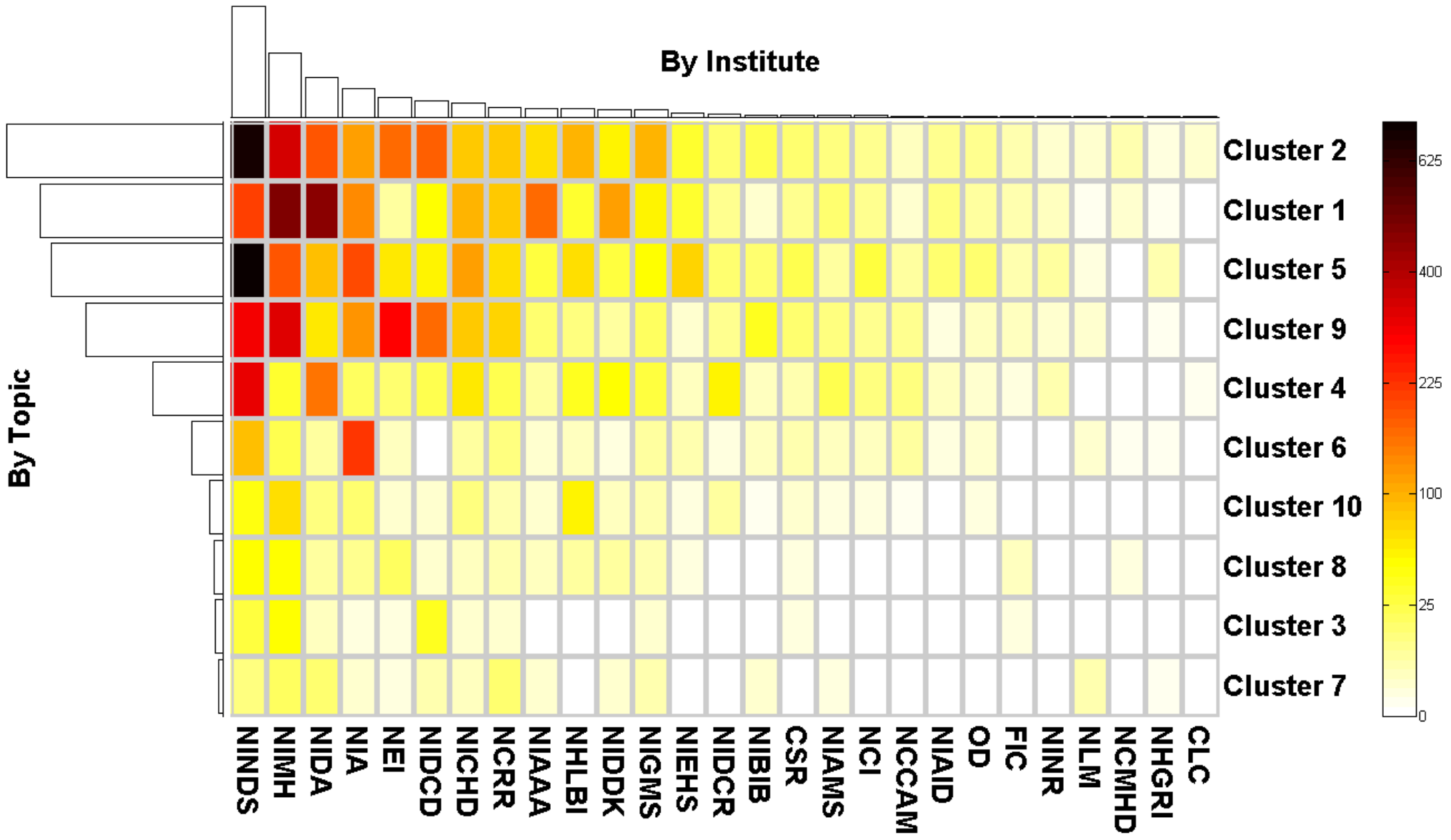

(c)

**Figure 12.** (a) Distribution of NIH funding across institutes and themes for the 2006 meeting abstracts. (b) Distribution of NIH funding across institutes and Theme F subthemes for the 2006 meeting abstracts. (c) Distribution of NIH funding across institutes and topic clusters. The color of an individual entry in the "image grid" indicates the number of abstracts from a particular theme (for (a)), subtheme (for (b)), or topic cluster (for (c), as determined by graph partitioning) that were funded by a particular NIH institute. Colors are scaled non-linearly for greater contrast. The "bar plots" on each axis indicate the total number of abstracts funded by a particular institute (top) or contained in a particular topic group (left). Both rows and columns have been sorted by total number of abstracts.

## 4.5 Related Work – Topic Modeling Analysis in Computer Science.

Several fields within computational linguistics (CL) use topic modeling, clustering and large-scale visualization efforts to analyze text corpora of varying degrees of size. Typically these text collections are non-scientific (either using sources such as Wikipedia with over 2 million pages, large-scale crawls of the world-wide-web or newstext). The National Library of Medicine's MEDLINE corpus is the standard data of choice for biomedical text mining (Kim and Tsujii, 2006). MEDLINE contains roughly 16 million documents and requires large-scale supercomputing methods to analyze using these methods (Newman et al., 2006, Personal Communication).

A number of techniques provide an alternative methodology to LSA for the analysis of topics and topic signatures (the associations between words within clusters) within text, these include the log-likelihood ratio (Lin and Hovy, 2000), a variety of clustering methods (See Pantel and Lin, 2002 for one example), and Latent Dirichlet Allocation (LDA, Blei et al., 2003). One refinement of LDA uses Gibbs sampling as an efficient methodology to discover topics (Griffiths and Steyvers, 2004; Newman et al., 2006). The complexity of the data may be explored with advanced graph visualization techniques to assist the analysis (Shiffrin and Borner, 2004). Recent studies include analyses of the 20 years of abstracts from the Proceedings of the National Academy of Sciences (PNAS, Boyack, 2004), or from publications concerned with Melanoma research (Boyack et al., 2004).

Unlike massive resources such as MEDLINE, the SFN annual meeting abstract data provides an ideal 'laboratory' for the use of these techniques on a small, focused set in the service of a small specific community. As a well-established method to investigate topics for our specific domain, we focused on the use of LSA to provide a clear high-level overview of the whole subject and to investigate detailed trends and issues concerning policy and the informational needs of neuroscientists. We envisage that the SFN abstracts can provide a valuable resource and application domain for the CL community since neuroscientists need efficient computational tools to assist them in their scholarly work

## MATERIALS AND METHODS

### Sources

The annual Society for Neuroscience (SFN) meeting abstracts from the years 2001 through 2006 were available as XML files on CD-ROMs from the SFN during the annual meetings of the society. These XML files were parsed to extract tagged attributes associated with each abstract. Each of these attributes was further processed in order to extract specific types of data. For example, the XML files provide attributes corresponding to authors' full names; these attributes were tokenized in order to separate last name from first and middle initials. Similar processing was applied to institution affiliations in which department name, institution name, city, state (for US and Canada), and country are identified. Furthermore, each author was linked to her respective institution based on annotated superscript numbers supplied during abstract submission. The postprocessed data were added into persistent storage in a MySQL database. The database contains three entity tables: author, institution, and paper. Since each author can be affiliated with multiple institutions and can produce one or more papers, these entities are mapped using many-to-many relationships in the database. For this study, we created one database for each year between 2001 and 2006, as well as a consolidated database encompassing data from all 6 years.

### Author Disambiguation

As is the case in many bibliographical resources, each author in an SFN abstract is identified by last name followed by one or more initials. Such an identification system is inherently ambiguous and can impact the quality of the database as more abstracts are pooled from multiple years. Two types of name ambiguities are observed during the parsing process. The first type results from the same author using a different number of initials in different abstracts. For example, Partha Mitra from Cold Spring Harbor Laboratory has been identified as "Mitra, P." and "Mitra, P. P." in different abstracts. Because such inconsistencies could lead to falsely identifying the same author as two unique individuals, only the last name and first initial were compared by default. Middle initials were used if and only if the two author names being compared both contained a middle initial. The second type of ambiguity arises when different authors actually share the same name and initials (e.g. "Brown, S." from the University of Tennessee in Memphis and "Brown, S." from Columbia University). To resolve this scenario, authors were identified as different individuals if their affiliations were different, regardless of name identities. This heuristic, of course, assumes that no two authors sharing the same name work in the same department of an institution, which is reasonable given the nature and size of the SFN data.

The method employed to distinguish authors by straightforward comparison of institution strings inevitably results in a large number of duplicates. This is because institution entities usually have many name variants. Syntactic differences ("Memorial Sloan Kettering Cancer Center" and "Sloan Kettering Institute for Cancer Research"), the use of abbreviations or acronyms ("New York State Psychiatric Institute" and "NYS Psychiatric Institute"), and even misspellings ("University of Pittsburg" instead of "University of Pittsburgh", and "Wilfred Laurier University" in Ontario, Canada instead of "Wilfrid Laurier University") were present due to the lack of a controlled vocabulary in abstract submission. Given this situation, a strategy that relies on exact string matching might suffer from low recall (Cohen and Sarawagi, 2004). This problem of determining whether different names refer to the same entity, or *entity matching*, has been addressed extensively in the field of information integration, and numerous solutions have been developed (Shen et al., 2007). Here, the following procedure was used to resolve semantic ambiguities for institution entities:

1. Break all institution affiliations, which consist of department name, institution name, city, state (for US and Canada), and country, into "bags of words" (or tokens). Convert all words to upper case.

2. Remove "stop words" from the token sets. Stop words are words that do not carry any weight in distinguishing different named entities. The initial stop list was downloaded from the Cornell SMART project (ftp://ftp.cs.cornell.edu/pub/smart/english.stop), and was supplemented by institution specific stop words such as "college", "clinic", "center", "laboratory", "program", "campus", etc.

3. Perform token based name matching using Jaccard similarity, which is defined as:

$$J = \frac{|S \cap T|}{|S \cup T|}$$

   where $S$ and $T$ are token sets of two arbitrary strings $s$ and $t$, respectively. Two institutions are considered identical if their Jaccard similarity is 1. This step resolves institution names with different word orders such as "Weill Medical College of Cornell University" and "Cornell University Weill Medical College".

4. Edit distance is used as a metric to resolve syntactic variations in institution names (e.g. "UC Berkeley" versus "University California Berkeley", or "Mount Sinai" versus "Mt. Sinai"). The edit distance between strings $s$ and $t$ is the cost of the best sequence of edit operations that convert $s$ to $t$ (Bilenko et al., 2003). If the distance between two names is less than a certain threshold, the two are considered aliases of the same entity and are thus merged into one representation.

In addition to institution entities, *co-authorship patterns* were also used to detect authors who moved between affiliations, further reducing duplicate author instances. For simplicity, authors who share the same name and have at least one common co-author were considered to be the same individual. The workflow for disambiguating and matching author entities is summarized in Figure 13.

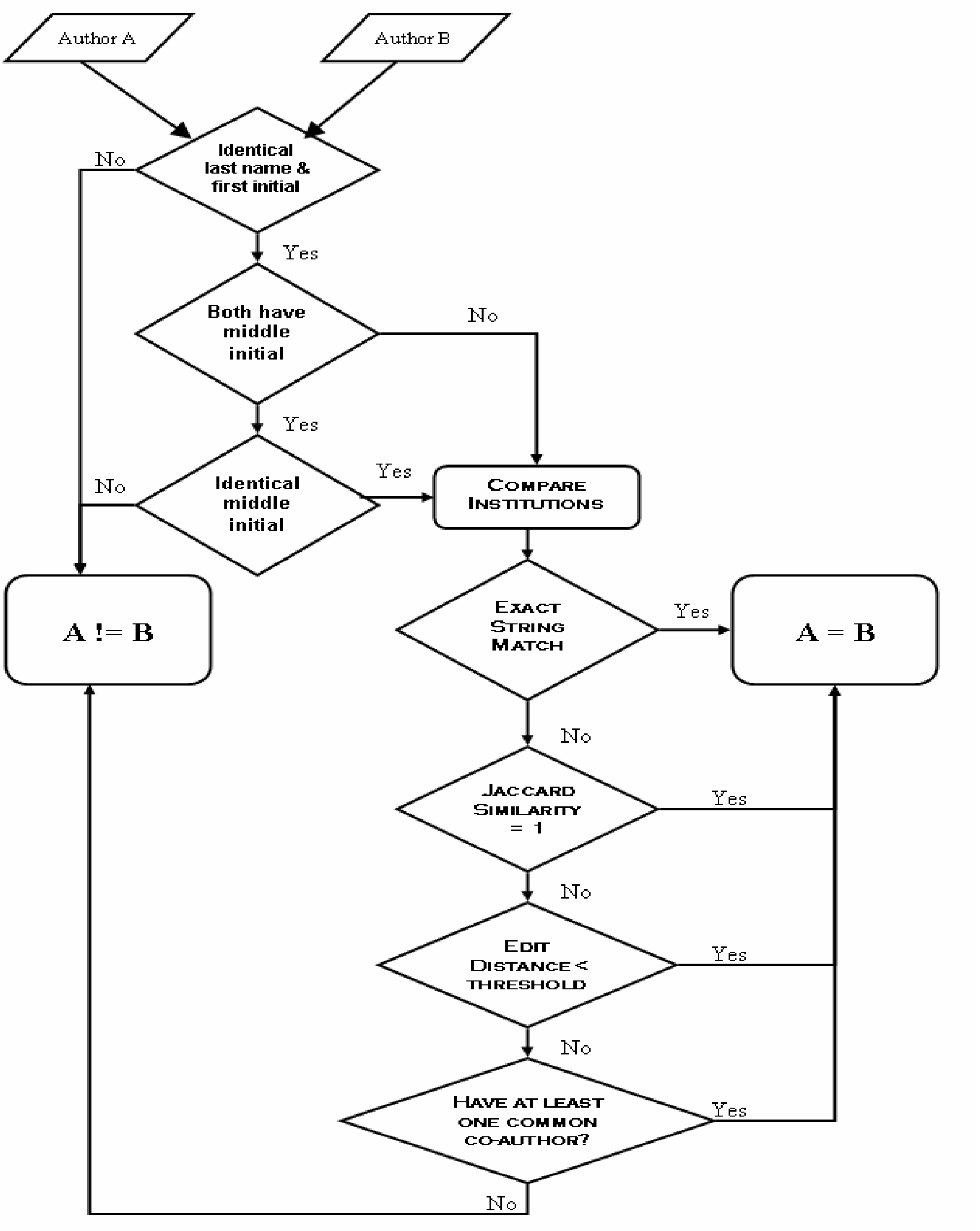

**Figure 13.** Work flow of determining whether two authors are the same individual.

## Graph Analysis

The breadth-first search algorithm used to calculate lengths of shortest paths between all pairs of authors for whom a connection exists was implemented in Perl. Refer to *Introduction to Algorithms* (Cormen et al., 2001) for detailed descriptions of the algorithm.

All other graph analysis (connected component, betweenness centrality, and clustering coefficient) were performed in MATLAB 7.3.0 (R2006b) using the MATLAB Boost Graph Library (MatlabBGL) written by David Gleich (http://www.stanford.edu/~dgleich/programs/matlab_bgl).

## Topic Modeling

### Latent Semantic Analysis

The first step of latent semantic analysis (LSA) was to construct a term-by-document matrix, $A$, in which each row corresponds to a unique term and each column to a unique document (abstract). Entry $A_{ij}$ contains the number of times term $i$ appeared in abstract $j$. The full text from the 87543 SFN meeting abstracts from years 2001 to 2006 were first parsed into tokens. All punctuations, numbers, and other special characters were discarded. In addition, common English words that do not carry semantic value were eliminated based on a "stop word" list from the Cornell SMART project (ftp://ftp.cs.cornell.edu/pub/smart/english.stop). To further reduce the size of the resulting "bag of words", all terms that appeared in only one abstract were eliminated. Word stemming algorithms from Snowball (http://snowball.tartarus.org) were also applied to all tokens so that morphologically similar words sharing the same root (e.g. "neuron", "neurons", "neuronal") were collapsed into one ("neuron"). Previous studies have indicated that the use of stemming can result in some improvement of the precision and recall of information retrieval (Hull, 1996).

The preprocessing steps resulted in 87543 documents and 35943 terms. The term-by-document matrix $A$ was constructed by counting the number of occurrences of each term in each document. In LSA, it is customary to transform this frequency matrix by some weight function to give better interrelations between term and document. In this work, the matrix $A$ was weighted using the log entropy function (Berry and Browne, 1999). The log entropy weight of each term $i$ is the product of its local weight $l_{ij}$ and global weight $g_i$ computed as

$$l_{ij} = \log_2(1 + A_{ij})$$

$$g_i = 1 + \left( \frac{\sum_j [p_{ij} \log_2(p_{ij})]}{\log_2 n} \right)$$

$$p_{ij} = \frac{A_{ij}}{\sum_j A_{ij}}$$

where $A_{ij}$ is the frequency of the $i$th term in the $j$th document, $p_{ij}$ is the probability of the $i$th term occurring in the $j$th document, and $n$ is the total number of documents in the corpus. The weighted frequency of each element from $A$ is then calculated by multiplying its local component by its global component. In other words, the *weighted* $m \times n$ term-by-document matrix, $F$, is defined as

$$F = (f_{ij}),\ where\ \ f_{ij} = l_{ij} \times g_i$$

The goal of using a weighting scheme is to assign less weight to terms that appear in many documents while awarding more weight to less frequent terms because the latter presumably have more differentiating power.

The weighted $m \times n$ term-by-document matrix, $F$, was factored into the product of 3 matrices using the singular value decomposition (SVD):

$$F = U \Sigma W^T$$

where $U$ is the $m \times r$ orthogonal matrix containing the left (term) singular vectors, $W^T$ is the $r \times n$ orthogonal matrix containing the right (document) singular vectors, and $\Sigma$ is the $r \times r$ diagonal marix of singular values of A (Golub and Van Loan, 1996). The number of singular values computed for the matrix $F$, denoted by $r$, was set to 100 in this work.

In the reduced dimensionality vector space created by truncating the SVD, terms that occur in similar documents are located near one another even if they never co-occur in the same document. Topically related documents are also grouped near one another in the reduced vector spaces. The similarity between any pair of documents $x$ and $y$ can be measured by their *cosine similarity*, which is computed as:

$$\cos(x, y) = \frac{x \bullet y}{|x||y|}$$

where $\boldsymbol{x}$ and $\boldsymbol{y}$ are the $r$-dimensional projections of the two documents in the reduced space.

**Topic Clustering**

After LSA was completed, topic clustering of the documents proceeded as follows. First, cosine similarities were computed exhaustively for all pairs of documents. For each document, a sorted list of nearest neighbors was identified as those having the highest cosine similarity scores. To reduce computational complexity, we identified only the top 100 nearest neighbors. Next, these data were represented as an undirected, weighted graph $G = (V, E)$ where each vertex, $v \in V$, denotes a document and each edge, $e(i, j) \in E$, connects a document $i$ with one of its nearest neighbors $j$, $i \neq j$. The weight associated with each edge $e(i, j)$ was simply set to $cos(i, j)$. Given the resulting sparse, connected graph, clustering could be performed using graph partitioning algorithms that segment the vertices of a graph into $n$ disjoint sets, $V_1, V_2, \ldots, V_n$, such that document similarity is high within a set $V_i$ and lower across different sets $V_i$ and $V_j$.

In this study, we applied the *Normalized Cuts* (NCuts) algorithm originally proposed by Shi and Malik (1994) to partition the full nearest neighbors graph. Unlike other graph partitioning methods, the NCuts algorithm avoids the bias of separating out small sets of isolated points by considering the global properties of the graph instead of focusing on local features (Shi and Malik, 1997). The algorithm attempts to partition $G$ into $n$ set of disjoint clusters by minimizing the normalized cut cost between any two partitions $V_i$, $V_j$, $V_i \cup V_j = V$ , $V_i \cap V_j = \varnothing$:

$$Ncut(V_i, V_j) = \frac{cut(V_i, V_j)}{assoc(V_i, V)} + \frac{cut(V_i, V_j)}{assoc(V_j, V)}$$

where $cut(V_i, V_j) = \sum_{u \in Vi, v \in Vj} w(u, v)$ is the sum of the weights of the edges that are removed between $V_i$ and $V_j$, $assoc(V_i, V) = \sum_{u \in Vi, t \in V} w(u, t)$ is the sum of the weights of edges connecting vertices in $V_i$ to all vertices in the graph, and $assoc(V_j, V)$ is similarly defined. Therefore, the NCuts algorithm not only evaluates the value of total edge weight connecting two partitions, but also computes the cut cost as a fraction of the total edge connections to all vertices in the graph (Shi and Malik, 1997) in order to produce globally optimal partitions. NCuts was applied recursively to cut the full graph into *n* connected components; the number of components or "clusters" is a parameter that we should estimate objectively.

**Estimate Number of Clusters**

Since the SFN theme labels and assignments were produced by scientists with domain expertise, we used this categorization as an evaluation benchmark to estimate the optimal number of clusters, *n*. The goal was to find the clustering of abstracts based on the NCuts algorithm that best matched the clustering based on SFN theme labels for the year 2006; this value *n* could then be assumed to be an appropriate number of clusters across the full 6-year data set. By varying the number of clusters, *n*, different degree of cluster agreement were obtained. We used the Adjusted Rand Index to quantify the agreement between NCuts clustering and the SFN theme labels. Adjusted Rand Index is defined as follows (Handl et al., 2005): Given two partitions *X* and *Y* of a common set of data points, the quantities *a*, *b*, *c*, and *d* are computed for all possible pairs of data points *i* and *j*, and their respective cluster assignments, $C_{X(i)}$, $C_{X(j)}$, $C_{Y(i)}$, $C_{Y(j)}$, where

$$a = |\{i, j \mid c_{X(i)} = c_{X(j)} \wedge c_{Y(i)} = c_{Y(j)}\}|$$
$$b = |\{i, j \mid c_{X(i)} = c_{X(j)} \wedge c_{Y(i)} \neq c_{Y(j)}\}|$$
$$c = |\{i, j \mid c_{X(i)} \neq c_{X(j)} \wedge c_{Y(i)} = c_{Y(j)}\}|$$
$$d = |\{i, j \mid c_{X(i)} \neq c_{X(j)} \wedge c_{Y(i)} \neq c_{Y(j)}\}|$$

In the present context, *X* represents SFN theme labels and *Y* represents the Ncuts clustering. The quantity *a* is the number of document pairs from the same SFN theme that are assigned to the same cluster in *Y*, *d* is the number of document pairs from *different* themes that are assigned to different clusters, *b* is the number of document pairs from the same theme that are assigned to different clusters, and *c* is the number of document pairs from different themes that are assigned to the same cluster.

The Rand Index (Rand, 1971) is then the fraction of all document pairs for which the clusterings agree:

$$R(X,Y) = \frac{a+d}{a+b+c+d}$$

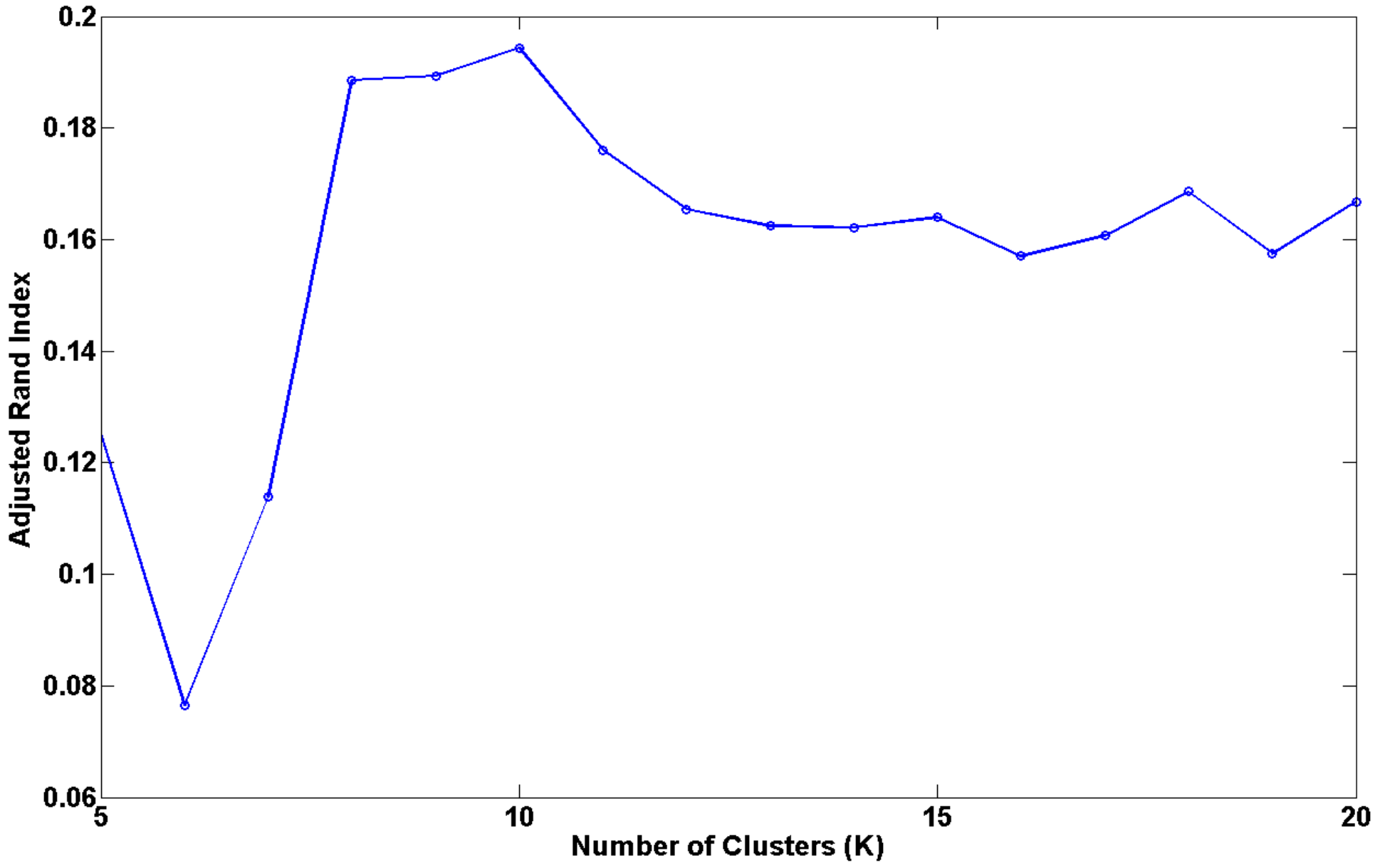

**Figure 14.** Adjusted Rand Index comparing similarity of SFN themes and NCuts-based partitioning versus number of clusters.

The Rand Index lies between 0 and 1. When the partitions *X* and *Y* agree perfectly, the Rand Index is 1. The Adjusted Rand Index was devised by Hubert and Arabie (1985) to correct for the fact that the expected value of *R* for random partitions is not constant. The Adjusted Rand Index linearly transforms the Rand Index such that its expected value is 0, and maximum value is 1. The Adjusted Rand Index comparing NCuts clusters with SFN themes was calculated for $n$=[5..20], a range intentionally chosen to be similar to the number of distinct SFN themes.

The Adjusted Rand Index versus the numbers of Ncuts clusters is shown in Figure 14. The plot suggests that Ncuts produces the clustering that is most similar to the SFN theme categorization when the number of clusters is 10, which was used throughout this work.

**Dynamics of Topics**

A 10 x 6 matrix, *D*, was constructed, where each element $D_{ij}$ denotes the number of abstracts from cluster *i* and year *j*. The matrix columns were normalized by the total number of abstracts in each year. To find topic clusters that demonstrate consistent and noteworthy rise or decline in popularity, we applied linear regression fit to the normalized frequency of each cluster by year.

An additional analysis of dynamics was performed using a term-frequency by year matrix, *F*. Entries of *F* count the occurrences of each term in abstracts, normalized by the total number of words in *all* abstracts for each year. Only those terms that appeared in more than one abstract were included in *F*. The row-wise mean, which indicates the average frequency of a given term *across years*, was removed. The singular value decomposition of this matrix was performed to reveal the principal temporal components and associated term-space components of change in the six year data set.

## Acknowledgements

This work was supported in part by a grant (#061984) from the W. M. Keck Foundation and the Crick-Clay Professorship from Cold Spring Harbor Laboratory.